\documentclass[10pt,letterpaper,final,twocolumn,journal]{IEEEtran}
\usepackage{setspace}
\usepackage{nomencl}
\makenomenclature
\usepackage{graphicx}

\usepackage{soul}

\usepackage{algorithm}
\usepackage{algorithmic}
\usepackage{epstopdf}
\usepackage{epsfig}
\usepackage[cmex10]{amsmath}
\usepackage{cite}
\usepackage{amssymb}
\usepackage[usenames,dvipsnames]{color} 
\usepackage{multirow}
\usepackage{array}
\long\def\symbolfootnote[#1]#2{\begingroup\def\thefootnote{\fnsymbol{footnote}}\footnote[#1]{#2}\endgroup}
\usepackage[table]{xcolor}
\usepackage{tabularx}
\usepackage{chngcntr}
\usepackage{booktabs}
\usepackage{caption}
\usepackage{subcaption}
\usepackage{float}
\usepackage{braket}

\usepackage{hyperref}

\usepackage{amsthm}
\theoremstyle{definition}

\newcommand{\subparagraph}{}

\usepackage{braket}

\begin{document}

\markboth{IEEE Access}{Accepted paper}

\title{{Experimental Characterization of Fault-Tolerant Circuits in Small-Scale Quantum Processors}}
\author{Rosie~Cane, Daryus~Chandra, Soon~Xin~Ng, Lajos~Hanzo, 
	\thanks{The authors are with the School of Electronics and Computer Science, University of Southampton, Southampton, SO17 1BJ, United Kingdom. E-mail: \href{mailto:lh@ecs.soton.ac.uk}{lh@ecs.soton.ac.uk}.}
	\thanks{The authors would like to acknowledge the financial support of the Engineering and Physical Sciences Research Council projects EP/N004558/1, EP/P034284/1, EP/P034284/1, EP/P003990/1 (COALESCE), of the Royal Society's Global Challenges Research Fund Grant as well as of the European Research Council's Advanced Fellow Grant QuantCom. We acknowledge the use of IBM Quantum services for this work.}
}

\maketitle

\begin{abstract}
Experiments conducted on open-access cloud-based IBM Quantum devices are presented for characterizing their fault tolerance using $[4,2,2]$-encoded gate sequences. Up to 100 logical gates are activated in the \textit{Ibmq\_Bogota} and \textit{Ibmq\_Santiago} devices and we found that a $[4,2,2]$ code's logical gate set may be deemed fault-tolerant for gate sequences larger than 10 gates. However, certain circuits did not satisfy the fault tolerance criterion. In some cases the encoded-gate sequences show a high error rate that is lower bounded at $\approx 0.1$, whereby the error inherent in these circuits cannot be mitigated by classical post-selection. A comparison of the experimental results to a simple error model reveal that the dominant gate errors cannot be readily represented by the popular Pauli error model. Finally, it is most accurate to assess the fault tolerance criterion when the circuits tested are restricted to those that give rise to an output state with a low dimension.
\end{abstract}


\section{Introduction}
\label{Introduction}
In the last few years, there have been vast leaps in the practical realisation of quantum computers, with both academia and industry demonstrating a variety of advances in the control, quality and number of programmable qubits \cite{kjaergaard2020superconducting}. The ever-increasing activities of the field have produced devices processing in excess of 50 qubits, by companies such as Google \cite{arute2019quantum}, IBM \cite{IBMref}, Rigetti \cite{rigettiref} and Ionq \cite{ionqiref} along with many smaller devices that are publicly available to the general research community. As the field evolves, these devices will have the capability to implement fully-fledged quantum algorithms with longer gate sequences. Within this framework, a fault-tolerant Quantum Error Correction Code (QECC) is needed for mitigating the accumulated component errors in a large-scale quantum circuit. Therefore, characterizing QECC's in new hardware can shed light on the device fidelity that will lead to the realisation of large-scale quantum algorithms. 

For a QECC to be fault-tolerant the circuits used for encoding, decoding and error correction must not introduce more errors than the code can correct since the gates of these circuits are imperfect, a single qubit error of a gate can proliferate through subsequent two-qubit gates that may overwhelm the codes' error correction capability. Hence, an encoding circuit built from a large number of realistic noisy gates may introduce too many errors at the start of the computation, making it impossible to achieve an overall coded error rate improvement. Fault-tolerant circuit design aims for mitigating the fundamental component errors inherent in QECC's. Therefore, to assess if a QECC is a viable method of improving the fidelity of a quantum algorithm, the additional circuitry used for implementing the QECC must be designed to be fault-tolerant. 

There are a number of QECC's that may be suitable for constructing fault-tolerant gate sequences at the time of writing. The family of attractive short QECC's includes the 5-qubit code of \cite{laflamme1996perfect}, the 7-qubit Steane code \cite{steane1996error}, topological codes \cite{dennis2002topological, IEEEtopological} and the 9-qubit Shor code \cite{shor1995scheme}. Fault-tolerant versions of these codes are prevalent \cite{preskill1998fault, shor1996FT}, but since the quantum hardware is still in its infancy, the functionality and architecture of the devices limits the choice of which QECC scheme can be tested. Fault-tolerant versions of a QECC often require many more qubits than just those used to encode a logical qubit. These overheads increase rapidly due to the need for multiple error correction iterations and ancilla check measurements. These processes often require a device that has multiple re-initialised ancilla qubits, as well unique qubit connectivity and classically controlled quantum operations. Therefore, experimental demonstrations of a fully fault-tolerant QECC with all the necessary steps such as repeated error detection and correction are still in the early stages of testing. 

\subsection{Fault-Tolerant Quantum Experiments}
\begin{figure}[htp]
    \centering
    \includegraphics[width=0.45\textwidth]{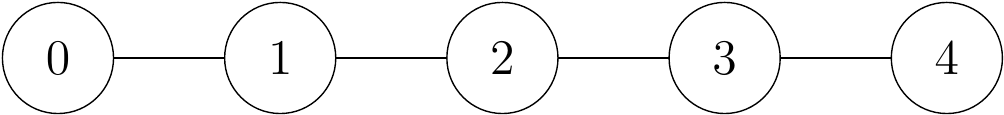}
    \caption{IBMQ 5-qubit \textit{Ibmq Santiago} device layout \cite{IBMref}.}
    \label{fig:fivequbit}
\end{figure}

\begin{figure}[htp]
    \centering
    \includegraphics[width=0.27\textwidth]{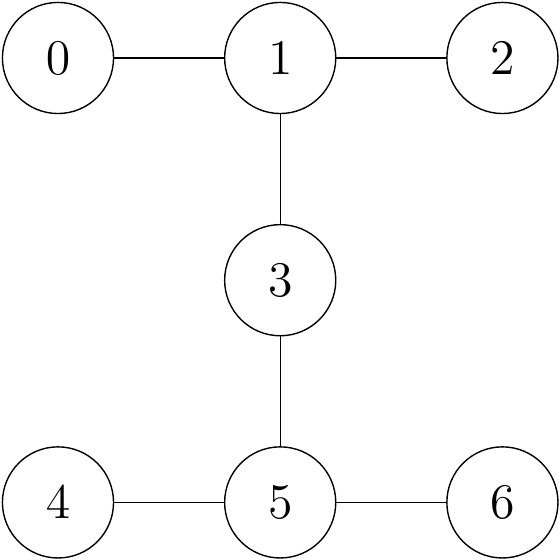}
    \caption{IBMQ 7-qubit \textit{Ibmq Casablanca} device layout \cite{IBMref}.}
    \label{fig:sevenqubit}
\end{figure}

There are a few considerations when selecting a quantum error correction code for experiments run on small-scale openly accessible state-of-the-art (SoA) devices. Figure~\ref{fig:fivequbit} and~\ref{fig:sevenqubit} show the device layout for the IBM Quantum (IBMQ) 5-qubit \textit{ibmq\_santiago} mode and for the 7-qubit \textit{ibmq\_casablanca} device, respectively \cite{IBMref}. These devices have a general qubit layout with certain two-qubit connections in a two-dimensional architecture. The general architecture of the device determines the possible placement of two-qubit gates in the physical circuit. For example, in Fig.~\ref{fig:fivequbit}, a Controlled-NOT (CNOT) gate may be placed between $q_{0} \rightarrow q_{1}$, but is not possible between $q_{0} \rightarrow q_{4}$. Therefore, in addition to having a sufficient number of available qubits, the circuit which applies the QECC in the proposed experiment must also be able to accommodate the specific qubit layout.

The limited functionality of SoA devices may also impose limitations on the extent to which error correction and detection can be applied. Methods that apply an error correction sub-routine typically require the measurement of a stabilizer or parity check operation to detect the presence of errors. Then the output of this measurement is successively forwarded to the necessary error correction regime. Therefore, to implement a typical measurement-based QECC the device must have the capability of carrying out a measurement during a particular computation and then input the result to a classically-controlled quantum gate. In addition, schemes that apply measurement-free error correction typically require a specifically-crafted device layout, whereby the code word qubits have nearest-neighbour connections with multiple ancilla qubits \cite{Crowthreshold}. Furthermore, this type of scheme is supported by ancilla qubits that can be reinitialized multiple times during a circuit's execution, so that the device architecture does not require an excessive number of connections to codeword qubits. Nevertheless, these features are theoretically realisable and there is rapid progress in expanding the capabilities of open-access devices underpinning the promise that a fully fault-tolerant implementation of a QECC may be possible in the near future. 

Given these limitations of the  existing hardware, the $[4,2,2]$ QECC is chosen for this study because of its straightforward implementation relying on only 4-5 qubits. This is because the fault-tolerant version only requires a single additional ancilla qubit \cite{shor1996FT}. Moreover, it is an error-detection code that relies on post-selection, which can be fulfilled in classical post-processing, circumventing multiple ancilla measurements in support of their circuit-based application. 
 
\subsection{State-of-the-Art Experiments}
Previous characterizations of the $[4,2,2]$ code have shown that the preparation of an encoded state and small-scale gate sequences offer an overall logical error rate improvement compared to its uncoded counterpart in the same device \cite{linke2017fault, takita2017experimental}. In \cite{takita2017experimental}, artificially inflicted errors were inserted, showing that indeed fault-tolerant circuit designs are robust to error proliferation. Vuillot demonstrated \cite{vuillot2017error} that small-scale encoded logical gates relying on error detection capability succeed in providing error-rate improvements, when the highest quality pair of qubits on the device are targeted. The comparison between non-fault-tolerant and the equivalent fault-tolerant circuits showed that the fault-tolerant design will have a lower logical error rate. It was also shown that the error rate of the circuit was influenced by the choice of the state sampled at the circuits output, observing that Pauli gate errors had less dramatic effect on the output distribution, when sampling from an equi-probable superposition of logical states.

Wilsch \textit{et. al.} compared various devices \cite{willsch2018testing}, showing that the fault tolerance criterion was only satisfied when certain types of underlying errors are present in the hardware, such as preparation and measurement errors of the IBM devices. However, the dominance of decoherence errors in the spin qubit device meant that it failed to demonstrate fault tolerance, despite applying a similar scheme. Further investigations in \cite{harper2019fault, kole2020resource} conclude that the overall performance improvement attained by the QECC coded scheme can be explained by the low circuit overheads involved in applying the most error-prone gates, namely two-qubit gates, in the logical code space. 

Against this backdrop, in this contribution we investigate the $[4,2,2]$-encoded gate sequences using the IBM Quantum services. Implementing small-scale QECC experiments in newly available devices provides a straightforward method of verifying that the QECC is constructed of fault-tolerant circuits according to the most realistic noise model in comparison to a result obtained by simulations. Moreover, this might allow us to asses a QECC's potential of enhancing a quantum algorithm without requiring large-scale classical simulations that meet a set of assumptions about the noise model. The results to be presented show that the $[4,2,2]$ code satisfies the fault tolerance criterion, because the uncoded scheme contains a larger number of two-qubit gates. However, we observe that the error rate of the coded scheme should still be significantly lower than what is observed and should also scale with the gate sequence length. Therefore, it is concluded that Pauli gate errors do not constitute the most important source of error in terms of quantifying the ultimate fidelity of the circuit. It will also be shown that post-selection may fail to fix certain proliferated qubit preparation errors. Furthermore, the encoding circuit may be very sensitive to those gate errors, which cannot be represented by the pure Pauli gate error model or to those that cannot be mitigated by post-selection\footnote{An over-rotation (or under-rotation) of the Hadamard gate in the encoding circuit may introduce an error that cannot be corrected in post-selection. Furthermore, the non-fault-tolerant encoding circuit implemented in this experiment may proliferate an error occurring during the preparation of the initialised qubit register. See Section~\ref{sec:exp1} for further discussion.}.

The structure of this paper is as follows. A fault tolerance criterion is defined for our experiments using SoA quantum processors in Section~\ref{sec:tft}. This is followed by Section~\ref{sec:design}, where our IBM-computer experiments are described. Then the $[4,2,2]$ code is presented in Section~\ref{sec:422Code}, along with circuit models as well as a method of extracting the relevant error rate metrics. In Section~\ref{sec:model} we define a simple Pauli-gate error model characterized by its gate and measurement error parameters. Finally, in Section~\ref{sec:ibmqresults} this model is compared to the experimental results of the $[4,2,2]$-encoded gate sequences, followed by our conclusions.

Against the aforementioned background, our novel contributions are: 
\begin{enumerate}
\item \textit{Using open-access IBMQ experiments, we show that the $[4,2,2]$ code's state preparation and its encoded logical gates satisfy a fault tolerance criterion for certain logical gate sequences, where the uncoded physical two-qubit gate count is lower than that of its coded counterpart.}
\item \textit{Our experimental results are compared to a simple error model having a small number of parameters for characterizing this QECC, which indicate the pivotal role of fault-tolerant designs in practical circuit construction.}
\item \textit{We observe that the QECC scheme is highly sensitive to errors close to the input of the circuit as well to qubit preparation errors that are proliferated by the encoding circuit. We demonstrate that the fidelity of the Hadamard gate used for initializing the encoded state will lower-bound the error rate performance of the coded scheme, when the CNOT gate error is mitigated by post-selection.}
\item \textit{Our results demonstrate that the trace distance measure only constitutes a reliable metric for certain QECC experiments, where the dimension of the ideal output is the same for all the sampled circuits. Stipulating this idealized experimental condition is necessary in order to maintain a consistent interpretation of the results. }
\end{enumerate}

\section{Experiment Design Using the [4,2,2] Code}
\subsection{Quantum Fault Tolerance Criterion}
\label{sec:tft}

In this section we consider (1) a traditional definition of fault-tolerant circuits \cite{N&C} and then (2) define a fault tolerance criterion more specifically suited to small-scale near-term experiments \cite{gottesman2016quantum}. 
 
\subsubsection{Fault Tolerant Circuit Design}
\label{sec:theoreticalFT}

Accordingly, we say that in a fault-tolerant circuit, an error from a single component will not overload the QECC, hence incurring zero logical errors after an error-correction step \cite{N&C, shor1996FT}. By contrast, a qubit error introduced by an individual gate of a non-fault-tolerant circuit can be proliferated to a larger number of errors by the application of noiseless successive gates\footnote{Note that this example and definition assumes a simple error model relying on interdependent component errors. Definitions relying on more practical assumptions can be found in \cite{aharonov1999fault, ng2009fault, aliferis2005quantum}.}. In other words, this has the effect of introducing an increased number of qubit errors into the circuit \cite{paper1}. For example, when a single bit flip error $XI$ is imposed on the control qubit of a CNOT gate, it will be copied to the target qubit and consequently the higher weight error of $XX$ will be output. Therefore, in a highly connected circuit having numerous independent qubits a single qubit error may overwhelm a QECC's error correction capability. More explicitly, a quantum circuit protected\footnote{An $[ n, k, d ]$  QECC encodes $k$ logical qubits into $n$ physical qubits. This has a minimum distance of $d$, and therefore is capable of correcting $e=[d-1/2]$ individual physical qubit errors \cite{N&C}.} by an $[n,k,d]$ QECC is only deemed fault-tolerant if a single component error results in less than $e=(d-1)/2$ individual qubit errors at the output of the circuit, where the code is capable of correcting $e$ errors upon using hard decisions.\\

\noindent \textbf{Definition 1}.A QECC is said to be fault-tolerant if an error occurring in a single circuit component results in either $e$ or less than $e$ individual qubit errors at the output of the circuit block \cite{N&C}. \\

This general definition can be verified either numerically or by simulation. For example, if a single component error occurs with probability $p$, the simulated error rate of the coded circuit block\footnote{Therefore, within this circuit block, two simultaneous component errors occur with probability $\mathcal{O}(p^{2})$, provided that the probability of a component error is independent.} is $p_{c}=\mathcal{O}(p^{2})$, provided that the probability of a component error is independent. This is because all single component errors occurring with a probability order of $\mathcal{O}(p)$ proliferate to qubit errors that can be corrected during an error correction step when the circuit design is fault-tolerant. However, eliminating the error from a single gate error will not remove circuit error entirely. The qubit error that cannot be corrected will occur from any configuration of two or more simultaneous gate errors. Nevertheless, a circuit that satisfies this definition of fault tolerance will exhibit an error rate improvement over the corresponding uncoded circuit, i.e $p_{c}<p_{u}$, because the uncoded scheme has an error rate with probability $\mathcal{O}(p)$. 

There are many fault-tolerant circuit designs satisfying \textit{Definition 1} \cite{N&C, campbell2017roads}. This framework has historically been verified using simulations, which rely on analytical error models that are assumed to imitate a real quantum processor. Using these models, diverse component error rate thresholds have been derived. Therefore, it has been shown that an arbitrarily long computation becomes possible, provided that the components operate below the maximum tolerable error rate \cite{knill1998resilient}. The value of this specific error rate threshold depends both on the noise model assumed, as well as on the particular choice of the fault-tolerant technique employed, and on whether the model has been determined analytically or by simulation. Simple  versions  of these  models  rely  on  an  unbiased  depolarizing  channel  suffering from independent  single-qubit  errors \cite{N&C} or from correlated errors using a general Hamiltonian framework \cite{aharonov1999fault, ng2009fault, aliferis2005quantum}. 

Given that several quantum processors are accessible in the cloud, it is possible to include experimental results in the process of developing fault-tolerant QECC's for characterizing device-specific errors. Naturally, it is desirable for the noise within the real device to be accurately characterised for the model obtained. This model may be limited to the parameters that define the most likely error patterns that occur in the device, hence making the calculations simple enough for an efficient classical simulation. The response of the fault-tolerant protocol to these parameters would also have to be known, and then the QECC can be specifically designed for the particular device considered. At this point a benchmark component error rate may be derived and bespoke hardware improvements can be recommended. 

However, we have to strike a trade-off between the complexity and accuracy of the classical simulation of the noise model. In this context, it is quite challenging to infer the error rate inflicted by an individual noise source in an interconnected circuit \cite{sanders2015bounding}. For example, repeated activation of a specific gate may incur its own nuanced interaction or there may be multiple ways a gate incurs an error. In some cases, the gate error rates determined using the randomized benchmarking technique of \cite{Randomized} may exclude certain types of gate-coherence errors\footnote{A coherence error can be thought of as something like a calibration error \cite{greenbaum2017modeling}. This is an over or under-rotation of the gate each time the gate is called. See Section~\ref{sec:exp1} and Appendix~\ref{sec:coherent}.}.

\subsubsection{Criterion for Small-Scale Experiments}
\label{sec:FTcriterion}

Therefore, the prediction of how a QECC will influence the estimated error phenomena rate occurring in a real device may be most straightforwardly carried out by a full experimental implementation of a QECC, which characterizes all sources of circuit errors. The methodology of \cite{gottesman2016quantum} provides a starting point for defining fault tolerance within the constraints of near-term devices. An experiment conducted using a prototype quantum processor may be said to demonstrate fault tolerance when: 
\begin{enumerate}
\item  the error rate of the encoded circuit $D_{c}$ is shown to be lower than that of its uncoded counterpart $D_{u} $;
\item it is a complete circuit implementation, which includes the initial state preparation and final measurement;
\item the output distribution of the encoded circuit is equivalent to that of the uncoded circuit;
\item both the encoded and uncoded experiments are run on the same device.
\end{enumerate}
For example, if the scheme satisfies $D_{c} <  D_{u}$, it is deemed to be fault-tolerant as long as the experimental assumptions (b)-(d) are upheld. The error rate of the circuit output $D_{u,c}$ is quantified in terms of the trace distance metric of the experimental outcome with respect to the ideal outcome defined in more detail in Section~\ref{sec:errorrate}.\\ 

\noindent \textbf{Definition 2} A QECC demonstrates fault tolerance in a small-scale quantum experiment when it satisfies $D_{c} <  D_{u}$ \cite{gottesman2016quantum}.\\

Directly characterizing fault-tolerant QECCs on an experimental quantum processor has a number of benefits. The results characterize the response of the QECC to the total noise model of the device \cite{iyer2018small}. Therefore, the experimental results incorporate both gate errors as well as qubit preparation and measurement errors, plus the effects of repeatedly activating components without any simplifying assumptions concerning the independence or correlation of error sources. The drawback of this however is that some coarse assumptions are required concerning the most likely source of error at the hardware level in order for the fault-tolerant protocol to be specifically optimised for the particular device at the software level. Nevertheless, the resultant benefit is that a specific characterisation of the device appears to be unnecessary for this approach. Therefore, the combination of experimental results with a simplified classical simulation may be the most convenient process of advancing the understanding of fault tolerance in QECCs for practical purposes. 

\subsection{Experiments Relying on Open Quantum Software}
\label{sec:design}
\begin{figure}[htb]
    \centering
    \includegraphics[width=0.45\textwidth]{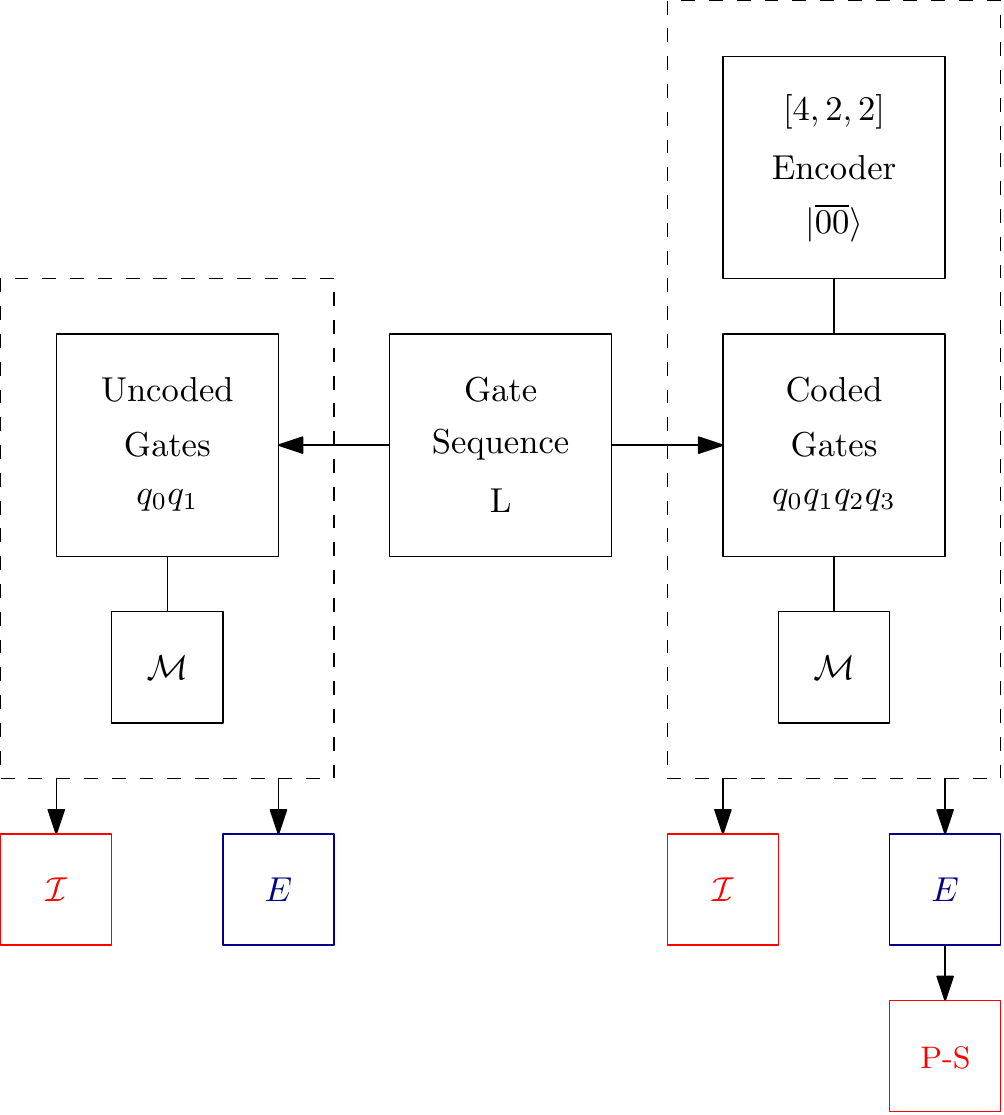}
    \caption{Schematic of the experiments design where $\mathcal{M}$ denotes a quantum measurement. Furthermore, the \textcolor{red}{\bf red} boxes indicate that the outcome is derived classically and the \textcolor{blue}{\bf blue} boxes represent a quantum experimental result. Finally, $\mathcal{I}$ and $E$ are the ideal and the experimental outcome distribution, respectively, while P-S denotes classical post-processing.}
    \label{fig:ibmexp}
\end{figure}

The results presented in this paper are obtained from the IBM Quantum cloud-based platform which provides access to a number of real quantum computers \cite{IBMref}. However, this methodology may also be applied to other quantum technologies relying on programmable circuits and multiple jobs per user. Figure~\ref{fig:ibmexp} represents a schematic of the methodology used for classifying a $[4,2,2]$-encoded circuit as fault-tolerant by comparing the uncoded and the corresponding encoded version of a gate sequence. 

The gate sequence length $L$ refers to the number of successive logical gates in the circuit that is being tested. For example, in the scenario of a QECC-protected quantum algorithm\footnote{For example, a real system that requires fault tolerant gate sequences may rely on any quantum algorithms associated with long gate sequences, such as Shors algorithm \cite{shor1999polynomial}.}, the sequence length $L$ would correspond to the number of gates in the circuit. The total circuit depth is given by the number of physical gates required for the implementation of the QECC applied to the qubit register plus that of the required logical gates. To elaborate, this encompasses both the encoding circuit as well as the physical gates that implement each individual logical gate. Therefore, the total physical gate count of the circuit corresponds to the circuit that is directly implemented in the hardware. 

\begin{figure}
 \centering
    \includegraphics[width=0.45\textwidth]{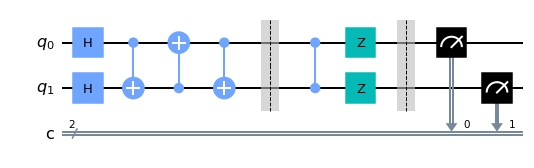}
    \caption[]
{Example of the uncoded version of a $L=2$ gate sequence according to Fig.~\ref{fig:ibmexp} in IBMQ \cite{IBMref}. This includes the logical gate sequence $H_{0} H_{1} \circ SWAP_{0,1}, \textrm{cz}_{0,1} \circ Z_{0}  Z_{1}$ and a final measurement.} 
    \label{fig:uncoded_ibm_fig}
\end{figure}

\begin{figure}
 \centering
       \includegraphics[width=0.45\textwidth]{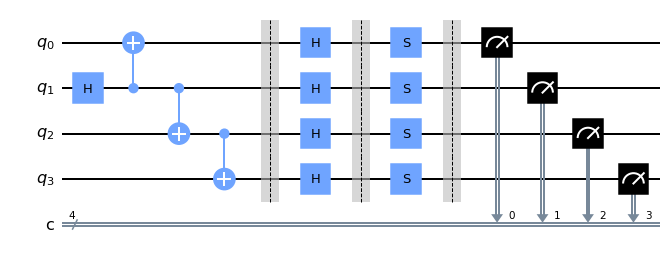}
    \caption[]
{Example of the coded version of the same $L=2$ gate sequence as Fig.~\ref{fig:uncoded_ibm_fig}. This includes the initial $|\overline{00}\rangle$ state preparation circuit (non-fault-tolerant version) followed by logical gate sequence $H_{0} H_{1} \circ SWAP_{0,1},\textrm{cz}_{0,1} \circ Z_{0}  Z_{1}$ and a final measurement.} 
   \label{fig:coded_ibm_fig}
\end{figure}

Accordingly, a unique gate sequence is generated for a sequence length $L$ and the corresponding physical circuit is then constructed for both the uncoded and encoded scheme. The encoded version includes both the encoding circuit and the final measurement (denoted by $\mathcal{M}$ in Fig.~\ref{fig:ibmexp}). The separate uncoded and encoded circuits are then implemented by the same hardware sequentially, represented by the pair of dashed boxes in Fig.~\ref{fig:ibmexp}. The uncoded circuit is realized both by a classical simulator (denoted $\mathcal{I}$) as well as by the quantum hardware (denoted $E$) to obtain the uncoded error rate $D_{u}$. Likewise, the encoded circuit is realised by both the classical and quantum hardware to obtain the coded error rate $D_{c}$. An example of the full circuit for an equivalent $L=2$ uncoded and coded gate sequence is shown in Fig.~\ref{fig:uncoded_ibm_fig} and Fig.~\ref{fig:coded_ibm_fig} respectively. The circuit shown in Fig.~\ref{fig:uncoded_ibm_fig} is represented by the dashed box at the left of Fig.~\ref{fig:ibmexp}. Likewise, Fig.~\ref{fig:coded_ibm_fig} is represented by the dashed box at the right of Fig.~\ref{fig:ibmexp}. 

The $L=2$ gate sequence in Fig.~\ref{fig:uncoded_ibm_fig} is the $H_{0} H_{1} \circ SWAP_{0,1}$ gate\footnote{The notation for the $H_{0} H_{1} \circ SWAP_{0,1}$ gate is written according to the circuit transformation (see Fig.~\ref{fig:H422}) rather than its mathematical form $SWAP_{0,1}[(H_{0}\otimes H_{1})|Q_{0}Q_{1}\rangle]$.} followed by the $\textrm{cz}_{0,1} \circ Z_{0}  Z_{1}$ gate (see Table~\ref{table:gates} in Section~\ref{sec:gates} for the definition). The uncoded circuit in Fig.~\ref{fig:uncoded_ibm_fig} shows each gate separated by circuit barrier, with a measurement at the end of the circuit (black squares). The corresponding encoded version of the circuit is shown in Fig.~\ref{fig:coded_ibm_fig}. The first section of the circuit is the encoding circuit (see Section~\ref{sec:422Code}). This is followed by the equivalent encoded version of the $H_{0} H_{1} \circ SWAP_{0,1}$ and $\textrm{cz}_{0,1} \circ Z_{0} Z_{1}$ gates. The circuit is then terminated with a measurement operation applied to each of the four physical qubits.

The general routine seen in Fig.~\ref{fig:ibmexp} is then applied to a selection of gate sequences, which form a sub-family that is representative of all possible gate sequences derived from the $[4,2,2]$-code's gate set. The complete set of gate sequences can be defined as all possible combinations of gates from a single one up to a length of $L$ gates. If the sub-family of typical gate sequences that are representative of the complete set satisfies the fault tolerance criterion, then it can be assumed that the QECC is fault-tolerant for any circuit \cite{gottesman2016quantum}. 

Both the uncoded and encoded circuits investigated rely on the same IBMQ device to keep the underlying source of error as similar as possible. The submission of the jobs to the IBMQ cloud-based platform is batched so that the encoded and uncoded versions are run one after another to our best ability. Each device is re-calibrated and all the jobs submitted for an experiment are within the same IBMQ calibration cycle. There are however user-specific restrictions both on the number of circuits and jobs according to the user access rights and the device chosen, as well as depending on the demand for the IBMQ device at a certain time. Note that despite batching results within a calibration cycle, the parameters governing the decoherence processes in superconducting qubits will vary with time\footnote{In addition, discussion of time-varying channel models in superconducting qubits can be found in \cite{martinez2021time}.} \cite{carroll2021dynamics, burnett2019decoherence}. Therefore, it is to be expected that there will be differences in results within each calibration cycle.

\subsection{Post-Selection}
\label{sec:PS}
The $[4,2,2]$ code of \cite{gottesman2016quantum} provides a method whereby a pair of logical qubits $Q_{0}Q_{1}$ are encoded using four physical qubits $q_{0}q_{1}q_{2}q_{3}$, as seen in Fig.~\ref{fig:coded_ibm_fig}. The $[4,2,2]$ code's logical states\footnote{Logical states and operations are denoted with an overbar; $\overline{x}$.} are \cite{gottesman2016quantum}:
\begin{align}
    |\overline{00}\rangle \rightarrow (|0000\rangle + |1111\rangle)/\sqrt{2} \label{eqn:00codeword} \\
    |\overline{01}\rangle \rightarrow (|1100\rangle + |0011\rangle)/\sqrt{2}  \label{eqn:01codeword}\\
    |\overline{10}\rangle \rightarrow (|1010\rangle + |0101\rangle)/\sqrt{2}   \label{eqn:10codeword}\\
    |\overline{11}\rangle \rightarrow (|0110\rangle + |1001\rangle)/\sqrt{2}  \label{eqn:11codeword}.
\end{align}

The $[4,2,2]$ code is an error detection code, therefore a codeword that is found to contain an error is discarded rather than corrected. To implement this scheme in small-scale experiments, error detection is carried out with the aid of classical post-processing. Explicitly, the logical qubits $Q_{0}Q_{1}$ can be measured in the computational basis by direct measurement of the physical qubit register $q_{0}q_{1}q_{2}q_{3}$. The operation of detecting an erroneous state by the classical post-processing is straightforward, because if the outcome of the measurement is a bit-string containing an even number of 1, then it can be decoded into one of the four legitimate codewords, of Eq.~\eqref{eqn:00codeword}-\eqref{eqn:11codeword}. If by contrast the measurement outcome corresponds to a bit-string with an odd parity; namely we have $1000, 0111,  0100, 1011, 0010, 1101,1110, 0001$, then an error has occurred, hence the corresponding results can be discarded in post-selection. Therefore, if $\gamma$ is the number of accepted legitimate results and $R$ is the total number of circuit outputs, then the post-selection retention ratio $r$ is defined by
\begin{equation}
    r = \frac{\gamma}{R},
    \label{eqn:ps_ratio}
\end{equation}
where $\gamma$ is equivalent to the number of outputs having even parity.

\subsection{[4,2,2]-Encoded State Preparation}
\label{sec:422Code}

\begin{figure}[htp]
    \centering
    \includegraphics[width=0.3\textwidth]{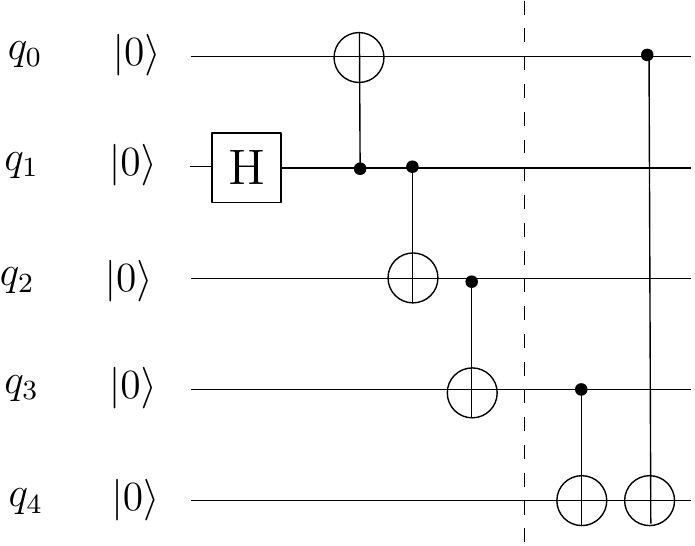}
    \caption{Fault-tolerant circuit to prepare the $[4,2,2]$-encoded logical state $|\overline{00}\rangle$. After the dashed line a parity check of $(q_{0},q_{3})$ is determined by measuring the ancilla in $q_{4}$. To make this circuit fault-tolerant post-selection is also applied to $q_{0}$ to $q_{3}$.}
    \label{fig:st_prep_422}
\end{figure}

There are several methods of ensuring that the logical state is encoded using a circuit relying on a theoretically fault-tolerant design, as presented in Section~\ref{sec:theoreticalFT}. To recap, if a single gate error proliferates through subsequent gates to an increased number of qubit errors that are not detectable according to the specific detection capability of the QECC, then the circuit design must not be deemed to be fault-tolerant \cite{N&C}. For example, if a certain CNOT gate in the $[4,2,2]$ code's encoding circuit has an erroneous output and this error in turn proliferates to an even number of qubit errors in the output state, then this error cannot be detected and the circuit is not fault-tolerant, as will be briefly exemplified below. Further discussions on error proliferation and fault-tolerant circuit design can be found for example in \cite{paper1,paper2}.
 
Figure~\ref{fig:st_prep_422} shows the circuit that prepares the $[4,2,2]$-encoded states $q_{0},q_{1},q_{2},q_{3}$ representing the logical state $|\overline{00}\rangle$ in Eq.~\eqref{eqn:00codeword}. This circuit has a fault-tolerant design for two reasons. Firstly, the gate errors that proliferate to an odd number of errors at the output of the encoding circuit will be discarded in the post-selection operation presented in Section~\ref{sec:PS}. For example, if the CNOT gate between $(q_{1}, q_{0})$ incurs an $XI$ Pauli error, the $X$ error on the control qubit will be proliferated by the following two CNOT gates between both $(q_{1}, q_{2})$ as well as $(q_{2}, q_{3})$. Hence, the output state $q_{0},q_{1},q_{2},q_{3}$ will be $(|0111\rangle + |1000\rangle)/\sqrt{2}$. Since this contains an odd number of errors in each 4-tuple, the state will be discarded during the classical post-selection. 

Secondly, there are some gate errors that proliferate to an even number of qubit errors and therefore cannot be detected by the classical post-selection. To detect these gate errors an additional parity check is appended between $(q_{0}, q_{3})$ using an ancilla qubit in location $q_{4}$. If the result is $1$ when the ancilla qubit is measured, it indicates that the intended encoded state $|\overline{00}\rangle$ has not been prepared and the run should be discarded. For example, an $IX$ due to a fault in the CNOT gate between $(q_{1}, q_{2})$ will produce the output state $(|0011\rangle + |1100\rangle)/\sqrt{2}$. This error will not be picked up during post-selection, since the number of errors is even and the state corresponds to $|\overline{01}\rangle$. However, it can be spotted by the ancilla measurement. Therefore, this ancilla measurement combined with classical post-selection would render the encoder fault-tolerant, according to \textit{Definition 1} of Section \ref{sec:theoreticalFT}. Note that the circuit in Fig.~\ref{fig:st_prep_422} prepares only the $|\overline{00}\rangle$ encoded state. See Appendix~\ref{sec:other_encoded} for circuits that directly prepare alternative encoded states.

\subsection{Encoded Gates}
\label{sec:gates}

In this section we will describe the method of protecting logical gates by the $[4,2,2]$ encoder. In classical communications the FEC encoded bits are modulated and may be corrupted by the channel at the output of the demodulator, which is then corrected by the FEC decoder. By contrast, in a quantum computer, the faulty logical gates inflict errors, which can be modelled by a quantum decoherence channel. To demonstrate an error detection-aided quantum computation process, rather than merely a protected quantum memory, it is necessary to apply the $[4,2,2]$ code to their logical gates. There exists a set of logical gates\footnote{For quantum gate definitions see \cite{N&C}.} whose error-free operation may be detected by the $[4,2,2]$ scheme. These logical gates carry out certain logical transformations between the four legitimate encoded states of Eq.~\eqref{eqn:00codeword}-\eqref{eqn:11codeword}. Each logical operation is implemented by a set of physical gates carrying out the desired logical transformation.

  \begin{figure}[ht]
  \begin{minipage}[b]{0.45\linewidth}
    \includegraphics[width=.9\linewidth]{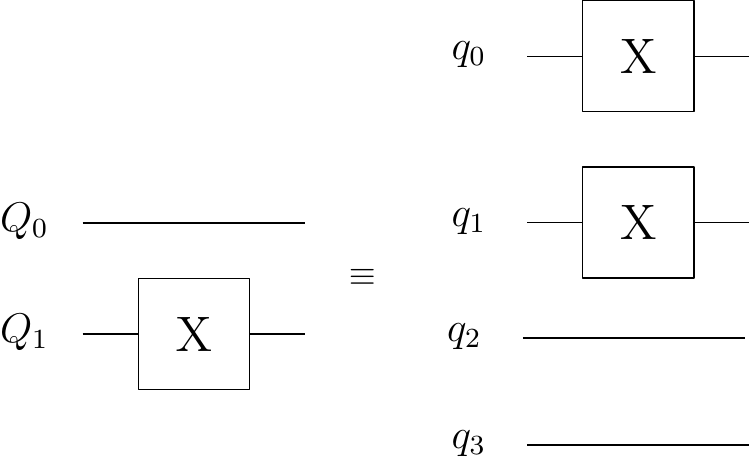} 
    \caption{$X_{1}$ Circuit} 
    \label{fig:x1422}
  \end{minipage} 
  \begin{minipage}[b]{0.45\linewidth}
    \includegraphics[width=.9\linewidth]{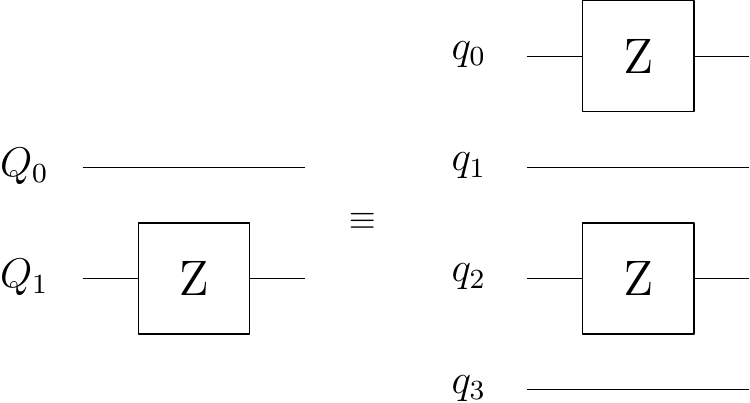} 
    \caption{$Z_{1}$ Circuit} 
    \label{fig:z1422}
  \end{minipage} 
  \end{figure}

The equivalent encoded and uncoded gate circuits are shown in Table~\ref{table:gates}. The $[4,2,2]$-encoded version of the gates has a physical circuit implementation that is fault-tolerant according to Section~\ref{sec:theoreticalFT} \cite{paper1}. For example, applying a logical $X$ gate to the $Q_{1}$ qubit in the encoded state $|\overline{00}\rangle$ is readily shown to be equivalent to applying two $X$ gates directly to the physical qubits $q_{0}$ and $q_{1}$ of the $[4,2,2]$-encoded state,\footnote{The logical gates may be applied after the fault-tolerant encoding scheme described in Section~\ref{sec:422Code}. Therefore the state $|\overline{00}\rangle$ is deemed error-free and the ensuing legitimate logical transformation applied to the code-word state will not result in error.} which is shown as follows
\begin{equation}
       \overline{X_{1}}|\overline{00}\rangle \rightarrow X_{0}X_{1}(\frac{1}{\sqrt{2}}\left( |0000\rangle+|1111\rangle) \right) = |\overline{01}\rangle,
\end{equation}
where $\overline{X_{1}}$ corresponds to the logical counterpart of an $X$ gate applied to the logical qubit $Q_{1}$. The circuit representing this gate is shown in Fig.~\ref{fig:x1422}. Additionally, the equivalent uncoded circuit is implemented by simply applying an $X$ gate to the uncoded qubit $q_{1}$. 

\begin{table}[htp]
\caption{Encoded and uncoded circuits according to the $[4,2,2]$ code logical gate set. For the definitions of quantum gates see \cite{N&C}.}
\centering
\begin{tabular}{ll}
\hline \hline
\textbf{Uncoded}      & \textbf{Coded}     \\ \hline \hline
$X_{0}$         & $X_{0}X_{2}$     \\
 $X_{1}$          & $X_{0} X_{1}$      \\
$Z_{0} $        & $Z_{0} Z_{1}$      \\
 $Z_{1} $        &$Z_{0} Z_{2}$     \\
$\textrm{cz}_{0,1} \circ Z_{0} Z_{1}$  & $S_{0} S_{1} S_{2} S_{3}$      \\
$H_{0} H_{1} \circ SWAP_{0,1}$ & $H_{0} H_{1} H_{2} H_{3} $     \\
% CNOT_{0,1}    & SWAP_{0,1} \\
\hline

\end{tabular}
\label{table:gates}
\end{table}

The circuit of the logical $Z_{1}$ gate is shown in Fig.~\ref{fig:z1422}. This is similar to the $X_{1}$ gate, but it is implemented by applying a $Z$ gate to qubits $q_{0}$ and $q_{2}$ in the $[4,2,2]$-encoded state. The logical single qubit gates $X$ and $Z$ have different physical gate implementations, depending on which qubit in the logical state is being targeted, as seen in Table~\ref{table:gates}. 

\begin{figure}[ht] 
  \begin{minipage}[b]{0.45\linewidth}
    \includegraphics[width=.9\linewidth]{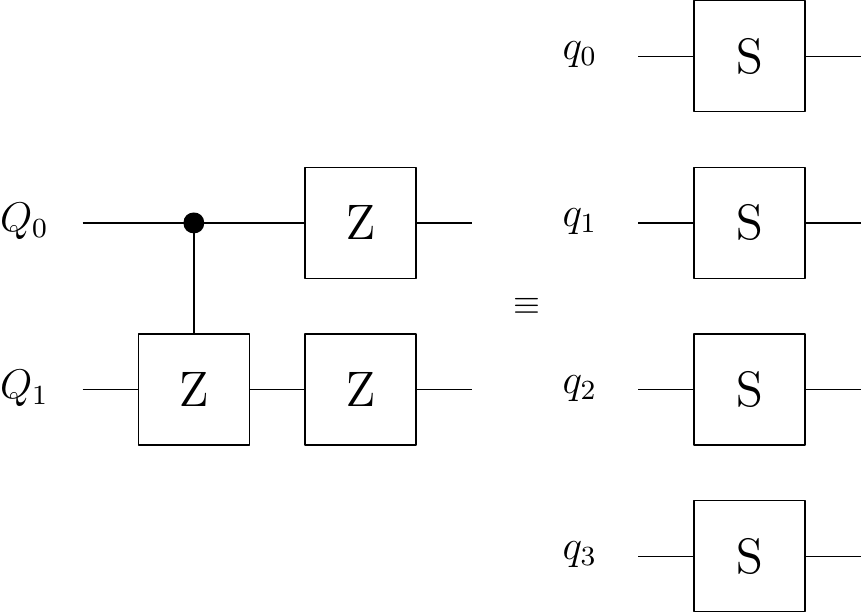} 
    \caption{$\textrm{cz}_{0,1} \circ Z_{0} Z_{1} $ Circuit} 
    \label{fig:cz422}
  \end{minipage} 
  \begin{minipage}[b]{0.45\linewidth}
    \includegraphics[width=.9\linewidth]{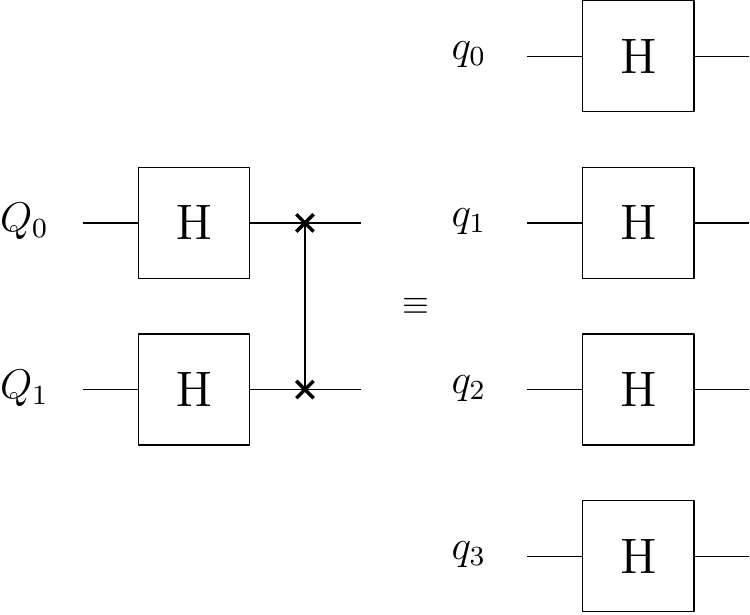} 
    \caption{$H_{0} H_{1} \circ SWAP_{0,1}$} 
    \label{fig:H422}
  \end{minipage} 
  \end{figure}

The gate referred to as $\textrm{cz}_{0,1} \circ Z_{0} Z_{1} $ is shown in Fig.~\ref{fig:cz422}. This gate is applied between $(Q_{0},Q_{1})$ and the uncoded version consists of a controlled-$Z$ (cz) gate followed by a $Z$ gate acting upon both qubits. The logically equivalent coded gate can be constructed by applying four of the single-qubit $S$ gates to the qubits $q_{0},q_{1}q_{2},q_{3}$. This has the advantage of implementing a logical two-qubit controlled-$Z$ gate by single-qubit gates\footnote{The $\textrm{cz}_{0,1}$ gate can be implemented alone by applying the $Z$ gates to the coded scheme with the physical gates $S_{0} S_{1} S_{2} S_{3} \circ Z_{1} Z_{2}$.} which has the effect of a low number of physical two-qubit gates in the encoded circuit.

Fig.~\ref{fig:H422} shows the effect of applying four physical Hadamard gates to the encoded state. This is the physical circuit that implements the logical gate referred to as $H_{0} H_{1} \circ SWAP_{0,1}$ in Table~\ref{table:gates}. This gate has the following transformation on the uncoded qubits $Q_{0},Q_{1}=|00\rangle$, giving the output:
\begin{align}
     \frac{1}{2}\Big( |00\rangle+|01\rangle+|10\rangle+|11\rangle \Big).
    \label{eqn:++}
\end{align}
The uncoded circuit of Fig.~\ref{fig:H422} is constituted by a pair of Hadamard gates $H_{0} H_{1}$ followed by a SWAP gate applied to $(Q_{0},Q_{1})$. The SWAP gate has the effect  of exchanging the position of two qubits $|xy\rangle \rightarrow |yx\rangle$ and is implemented using three CNOT gates \cite{N&C}. 

There are several ways of applying a CNOT gate to the qubits $(Q_{0},Q_{1})$ for the $[4,2,2]$ code \cite{gottesman2016quantum}. Applying a SWAP gate between qubits $(q_{0},q_{1})$ in the encoded state has the effect of a logical CNOT gate, but this is not a fault-tolerant circuit according to Section~\ref{sec:theoreticalFT}. A way around this is to apply a virtual SWAP gate between $(q_{0},q_{1})$ by switching the qubit positions in post-processing. Finally, with the aid of an additional ancilla qubit it is possible to use SWAP gates, but the excessive overheads of this circuit makes it less practical.

\section{Circuit Error Rate Evaluation}
\label{sec:errorrate}

The error rate of the circuit output is determined by quantifying the trace distance between the non-ideal experimental results and the ideal outcome distribution \cite{gottesman2016quantum}. This is the most practical metric that may be determined experimentally since it is operationally efficient, when the scale of the experiment is restricted by the number of available circuit activation's. The trace distance is obtained by measuring the final state at the circuit output in the computational basis. This gives a non-ideal noisy probability distribution, which is then compared to a classically simulated ideal distribution for the same circuit. Therefore, a low trace distance is desirable, since this corresponds to a circuit having a lower error rate. This procedure is repeated separately for both the encoded and equivalent uncoded scheme. This allows the error rates to be compared and the fault tolerance criterion $D_{c} <  D_{u}$ to be evaluated for that particular circuit, following \textit{Definition 2} of Section~\ref{sec:FTcriterion}.

Let $p$ be the ideal output distribution extracted from a classical circuit simulator and $\tilde{p}$ be the direct measurement outcome gleaned from the IBMQ device. For the uncoded scheme let the ideal probability distributions be denoted by $p^{u}$. This is the probability distribution over the set of possible outputs, when the qubits $Q_{0}Q_{1}$ are measured, namely $00,01,10,11$. The error-prone experimental circuit produces a different probability distribution $\tilde{p}^{u}$ over the same possible outcomes $00,01,10,11$. Then the error rate $D_{u}$ of the circuit output for the uncoded scheme is given by 
\begin{equation}
    D_{u} = \frac{1}{2} \sum_{i} |p^{u}_{i} - \tilde{p}^{u}_{i}|,
    \label{eqn:uncoded_D}
\end{equation}
where $i$ is the index of the set of possible outcomes. The error rate $D_{c}$ for the encoded scheme is given by the same method:
\begin{equation}
    D_{c} = \frac{1}{2} \sum_{i} |p^{c}_{i} - \tilde{p}^{c}_{i}|,
    \label{eqn:coded_D}
\end{equation}
where $p^{c}$ and $\tilde{p}^{c}$ are the ideal and non-ideal experimental results respectively, over the 16 possible outcomes for $q_{0}q_{1}q_{2}q_{3}$.

\section{Experimental Parameters}
\label{sec:model}

It is anticipated that two-qubit gates will have a more significant contribution to the overall error rate of the circuit, which is currently reflected in the benchmarked device metrics \cite{gambetta2012characterization}. Therefore, we seek to investigate whether the fault tolerance criterion is satisfied, because there is a larger number of two-qubit gates in the uncoded scheme \cite{harper2019fault, kole2020resource}. It will only become clear which the most critical parameters are after assessing the overall device noise effects. Nevertheless, in this section we assign dedicated parameters to the single gate error $\epsilon_{1}$ and to the two-qubit gate error $\epsilon_{2}$ as well as to the measurement error $P_{m}$ using a simple Pauli error model\footnote{This method is comparable to those numerical methods, where the gate errors are modelled independently by a Pauli error channel \cite{paper1, paper2}. Each gate error is treated as an independent error event, whereby the proliferation of qubit errors is mitigated by decoding. Therefore, each circuit block that is completed by an error correction step may be deemed fault tolerant.} \cite{N&C, preskill1998fault}. 

\subsection{Error Rate Associated with a Single Parameter}
\label{sec:single_param}
The associated measurement error can be accounted for by an independent qubit error channel having a single parameter. Let us consider the measurement error in a two-qubit register reminiscent of the uncoded scheme. Let us denote the probability of a single qubit read-out error by $0 < P_{m} < 1$. If the intended measurement outcome is $00$ but instead either $01$ or $10$ are measured, then we can say that a single qubit is measured incorrectly with probability $P_{m}(1-P_{m})$. Likewise, the measurement outcome is $11$ with probability $P_{m}^{2}$ since two qubits are simultaneously read out incorrectly. Then the total error rate becomes:
\begin{equation}
     \mathcal{E}_{M} = 1 - (1-P_{m})^{2}=2P_{m} - P_{m}^{2}.
     \label{eqn:uncoded_Pe}
\end{equation}

By the same reasoning, the corresponding encoded scheme will have an error rate according to the measurement of 4-qubit strings followed by the action of post-processing. All the odd numbers of qubit errors will be spotted and discarded by the post-selection. Therefore, $[4,2,2]$-encoded scheme incurs an error rate of $6P_{m}^{2}(1-P_{m})^{2}$ after post-selection according to the associated simultaneous two-qubit errors. 

\subsection{Encoder Gate Error}
\label{sec:encoder_error}

Unless the error in the encoding circuit can be perfectly corrected or the run  may be discarded, the error rate of the encoded scheme will be lower bounded by the residual encoder error. The ancilla measurement between $(q_{0},q_{3})$ of Fig.~\ref{fig:st_prep_422} does not have a straightforward implementation based on the device layouts shown in Fig.~\ref{fig:fivequbit} and Fig.~\ref{fig:sevenqubit}. Therefore, the ancilla measurement is excluded how the experiments presented in the next section. This means that the encoding circuit implemented does not have a fault tolerant design satisfying \textit{Definition 1} of Section~\ref{sec:theoreticalFT}. 

Let us assume that the error imposed by each gate may be modelled by a symmetrical Pauli error channel. A CNOT gate modelled by a two-qubit depolarizing channel outputs $IX,YI,YX,ZZ \dots$ after the normal functioning of the gate (see \cite{paper1, paper2} and Appendix~\ref{sec:depolarizing}). Each error has a probability of $\epsilon_{2}/15$, since there are $15$ combinations of $\{X,Y,Z,I\}$ excluding $II$ representing the identity operation that has no effect. The resultant gate error rate of the encoding circuit seen in Fig.~\ref{fig:st_prep_422} is $\mathcal{E}_{E} = \epsilon_{1}+3\epsilon_{2}$ before post-selection.

Let us consider the effect of each gate separately. Any $X,Y,Z$ error occurring after the Hadamard gate with probability $\epsilon_{1}/3$ will be proliferated by the following CNOT gates to a state with the outcome distribution of $|\overline{00}\rangle$ in Eq.~\eqref{eqn:00codeword}. Therefore, this error can be ignored. Let us assume that the first CNOT gate between $(q_{1},q_{0})$ of Fig.~\ref{fig:st_prep_422} has error probability of $\epsilon_{2}$. The phase flip errors $IZ,ZI,ZZ$ occurring with probability $3\epsilon_{2}/15$ can be ignored, since an odd-weight $Z$ error is not detectable in the $|\overline{00}\rangle$ state during post-selection and an even weight $Z$ error will cancel one another. In addition, all other depolarizing error combinations on this gate will result in an odd number of qubit errors, which will be discarded during post-selection or return the state to $|\overline{00}\rangle$. 

Therefore the final error rate of the encoding circuit will be determined by that of the CNOT gates connecting $(q_{1},q_{2})$ and $(q_{2},q_{3})$. Any of the $IX,IY,ZX,ZY$ errors after the $(q_{1},q_{2})$ CNOT gate will result in an even number of errors, namely in the $|0011\rangle+|1100\rangle$ state, therefore the error arising from this gate that cannot be detected occurs with a probability of $4\epsilon_{2}/15$. Any other error combinations applied to this gate will result in an odd number of errors that can be removed by post-selection. Likewise, the $(q_{2},q_{3})$ CNOT gate of Fig.~\ref{fig:st_prep_422} will also contribute $4\epsilon_{2}/15$ to the final error rate. Therefore, when considering gate errors modelled by the depolarizing channel it is expected that the encoding circuit will contribute $8\epsilon_{2}/15$, when the additional ancilla measurement is not implemented. Note that if the device layout is suitable for realizing the fully fault-tolerant circuit of Fig.~\ref{fig:st_prep_422} (which includes the ancilla parity check), then theoretically all possible gate errors occurring in the circuit are detectable and the above lower bound would not be applicable.

\subsection{Circuit Gate Error}

\begin{table}[htp]
\caption{Gate error probabilities according to the physical gate count for the circuits of Table~\ref{table:gates}. }
\centering
\begin{tabular}{lll}
\hline \hline
\textbf{Gate}      & \textbf{Uncoded} &   $\mathbf{[4,2,2]}, r=1$   \\ \hline \hline
$X_{0}, X_{1}, Z_{0}, Z_{1}$         & $\epsilon_{1} $ & $2\epsilon_{1} + \epsilon_{1}^{2}$  \\
$\textrm{cz}_{0,1} \circ Z_{0} Z_{1}$   & $2\epsilon_{1} + \epsilon_{2} $ & $4\epsilon_{1} + 6\epsilon_{1}^{2}$     \\
$H_{0} H_{1} \circ SWAP_{0,1}$ & $2\epsilon_{1} + 3\epsilon_{2}$ & $4\epsilon_{1} + 6\epsilon_{1}^{2}$     \\
% $CNOT_{0,1}$    & $t\epsilon_{2}$ & $\epsilon_{1} +(3t+1)\epsilon_{2}$ \\
\hline
\end{tabular}
\label{table:epsilon}
\end{table}

Let us assume that the circuit is modelled by a sequence of temporally uncorrelated noisy channels and consists of spatially uncorrelated physical gates, where $\mathcal{P}$ denotes the overall error rate of each physical circuit block that implements a logical gate in the sequence. Furthermore, there are $L$ gates in the sequence, each having an error rate $\mathcal{P}$. According to these idealized simplifying assumptions, the overall error rate $\mathcal{E}_{\mathcal{P}}$ of the gate sequence is given by
\begin{equation}
    \mathcal{E}_{\mathcal{P}} = \sum_{i=1}^{L} \binom{L}{i} \mathcal{P}^{i} = 1-(1-\mathcal{P})^{L}
    \label{eqn:gate_error_model}
\end{equation}
and $L\mathcal{P}$ is the largest term corresponding to the probability of a single logical gate block in the sequence operating with an error. Let us denote the physical single-qubit gate count by $n_{1}$ and the two-qubit gate count by $n_{2}$. Furthermore, the average physical single-qubit gate error probability is denoted by $\epsilon_{1}$, regardless of the specific type of the individual gate applied. Likewise, the average two-qubit gate error probability is $\epsilon_{2}$. For example, a logical $Z_{0}$ gate is implemented using $n_{1}=2$ physical $Z$ gates, each having an error rate of $ \epsilon_{1}$. Then the total gate error probability attributed to each circuit block is $\mathcal{P} = \tilde{\epsilon}_{1} + \tilde{\epsilon}_{2} +\tilde{\epsilon}_{1}\tilde{\epsilon}_{2}$, where we have
\begin{equation}
    \tilde{\epsilon}_{1} = \sum_{i=1}^{n_{1}} \binom{n_{1}}{i} \epsilon_{1}^{i}, \quad  \tilde{\epsilon}_{2} = \sum_{i=1}^{n_{2}} \binom{n_{2}}{i} \epsilon_{2}^{i}.
    \label{eqn:epsilon_tilde}
\end{equation}

Table~\ref{table:epsilon} shows the expected error rate of a circuit block implementing both the uncoded as well as the $[4,2,2]$-encoded scheme and the post-selected coded scheme. For example, the $X_{0}$ gate contains a single $X$ gate for the uncoded implementation, therefore we have $\mathcal{P} = \epsilon_{1}$. The corresponding encoded version requires $n_{1} = 2$ physical $X$ gates. Before post-selection ($r=1$) this circuit block will have an error rate of $\tilde{\epsilon}_{1} =2\epsilon_{1} + \epsilon_{1}^{2}$. After post-selection ($r<1$) the odd numbers of qubit errors are removed, so it is expected that we have $\mathcal{P} = \epsilon_{1}^{2}$. Since this circuit contains only single qubit gates, no qubit errors may proliferate to a larger number of errors through two-qubit gates. Additionally, the gate counts are the same for the single qubit encoded gates and $\epsilon_{1}$ represents the error probability of all individual physical gates, so by the same reasoning as that for the $X_{0}$ gate, the expected error rate attributed to the implementation of the $X_{1},Z_{0},Z_{1}$ gates can be derived. 

\section{IBMQ Experimental Results Associated with a Simple Error Model}
\label{sec:ibmqresults}

In this section we introduce three experiments. Each experiment relies on random sequences from the $[4,2,2]$-encoded gate set; $\{X_{0},X_{1},Z_{0},Z_{1}, \textrm{cz}_{0,1} \circ Z_{0} Z_{1}, H_{0} H_{1} \circ SWAP_{0,1}\}$. The first experiment in Section~\ref{sec:exp1} shows the results of implementing Fig.~\ref{fig:ibmexp} for random sequences of a reduced gate set that excludes the $ H_{0} H_{1} \circ SWAP_{0,1}$ gate. In the second experiment, sequences of the $H_{0} H_{1} \circ SWAP_{0,1}$ gate alone are considered and the results are discussed. This gate prepares an output state that is 4-dimensional therefore there are some considerations when deriving the error rate compared to an ideal state. The final experiment in Section~\ref{sec:other} shows the results of random sequences of the full gate set. Before we discuss the experiment results, let us consider the trace distance bounds in a \textit{worst case} circuit noise scenario.

\subsection{Trace Distance Bounds}
\label{sec:upperbound}

Consider the scenario where the only source of circuit error is the depolarizing channel. In the \textit{worst case} scenario the probability of error is $\xi=1$ meaning that the experimental circuit output is always the totally mixed state~\cite{N&C} 
\begin{equation}
    \frac{\textrm{I}}{4} = \frac{1}{4}\sum_{i=1}^{4} |i\rangle \langle i|.
    \label{eqn:mixed}
\end{equation}
This can be thought of as a randomized output, where the desired state has been totally corrupted by circuit error. In this case, the uncoded experimental output distribution $\tilde{p}^{u}$ is of the form: 
\begin{equation}
    \tilde{p}^{u}_{j} = \frac{1}{4} \quad \forall \quad j = \{ 00, 01, 10, 11\},
    \label{eqn:uncoded_noisy}
\end{equation}
where each measurement outcome is equi-probable.

First let us compare this to the class of circuits, where the ideal circuit output is 1-dimensional, so $p^{u}_{i} = 1$ for any $i = \{ 00, 01, 10, 11\}$. The dimension of the output state is determined by the selected gate sequence, namely by the specific state which that particular set of gates gives rise to. 

For example, if the circuit prepares $Q_{0}Q_{1} \equiv |00\rangle$ and then applies the gate $X_{1}$, the output becomes:
\begin{equation}
  |00\rangle \xrightarrow{X_{1}} |01\rangle.
  \label{eqn:circuit_01}
\end{equation}
In this case the ideal noiseless output generated by the classical simulator is
\begin{equation}
    p^{u}_{01} = 1, \quad p^{u}_{i} = 0 \quad \forall \quad i = \{ 00, 10, 11\}.
    \label{eqn:describe_01}
\end{equation}

When the ideal circuit output is given by Eq.~\eqref{eqn:describe_01} but Eq.~\eqref{eqn:uncoded_noisy} is the measured experimental distribution, the error rate becomes 
\begin{equation}
    D_{u} = \frac{1}{2}(|p^{u}_{01}-\tilde{p}^{u}_{j}|+3|p^{u}_{i}-\tilde{p}^{u}_{j}|) = 0.75
    \end{equation}
by Eq.~\eqref{eqn:uncoded_D}.

Now, consider the logically equivalent scenario for the $[4,2,2]$-encoded scheme. This is the encoded equivalent to the circuit in Eq.~\eqref{eqn:circuit_01} and generates the output state $|\overline{01}\rangle = (|1100\rangle + |0011\rangle)/\sqrt{2}$ given in Eq.~\eqref{eqn:01codeword}. Therefore, the ideal output of the noiseless classical simulation is
\begin{gather}
    p^{c}_{1100,0011}= \frac{1}{2}, \quad p^{c}_{i}= 0 \nonumber \\ \forall \quad i = \{ 0000, 1111,  0101, 1010, 0110, 1001\}.
\end{gather}
When $\xi=1$, the experimental circuit output is of the state $\textrm{I}/16$. After post-selection, we have 
\begin{gather}
  \tilde{p}^{c}_{j}= \frac{1}{8} \quad \forall \quad  j = \{ 0000, 1111, 0101, 1010, \nonumber \\ 0110, 1001, 1100,0011\},
    \label{eqn:coded_noisy}
\end{gather}
where the probability of each legitimate codeword is identical and it is normalised by the post-selection ratio of $r=1/2$. Then according to Eq.~\eqref{eqn:coded_D} the upper bound for the error rate of the encoded scheme is the same as that of the uncoded version, namely
\begin{equation}
    D_{c} = \frac{1}{2}(2|p^{c}_{1100,0011}-\tilde{p}^{c}_{j}|+6|p^{c}_{i}-\tilde{p}^{c}_{j}|) = 0.75.
\end{equation}

\subsection{Experiment 1: Reduced Gate Set}
\label{sec:exp1}

\begin{figure}[htp]
    \centering
    \includegraphics[width=0.5\textwidth]{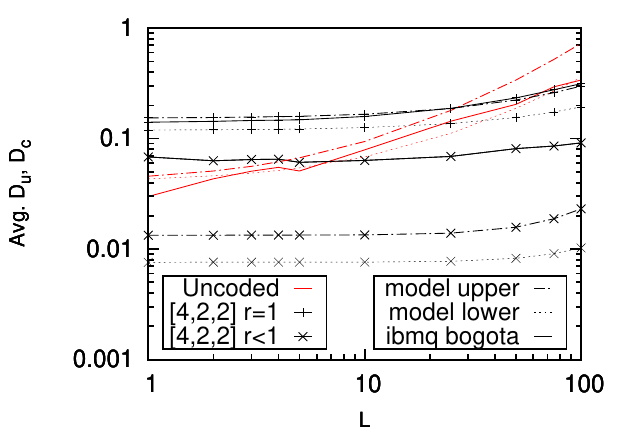}
    \caption{Experimental results based on the \textit{Ibmq\_Bogota} device characterizing random sequences of the $[4,2,2]$-encoded gates along with those of the corresponding uncoded gate for sequence lengths $L$. Model parameters: samples = 60, device = ibm bogota, date= 26.05.2021, gate set= $[X_{0},X_{1},Z_{0},Z_{1}, \textrm{cz}_{0,1} \circ Z_{0}  Z_{1}]$, $P_{m} = 0.02 $, $\epsilon_{1}:\epsilon_{2}$ = $1:40$ , $3 \times 10^{-3} < \epsilon_{1} < 5.5 \times 10^{-3}$ }
    \label{fig:bogota_RS_06}
\end{figure}

Figure~\ref{fig:bogota_RS_06} shows the results of random $[4,2,2]$-encoded gate sequences of length $1 \leq L \leq 100 $ after the initialisation of the $|\overline{00}\rangle$ encoded state, which were run on the \textit{Ibmq\_Bogota} device according to the method shown in Fig.~\ref{fig:ibmexp}. Let us compare this to a simple model having as few as three parameters; namely the single and two-qubit gate error as well as another parameter representing the measurement error defined in Section.\ref{sec:model}. In this section the results refer to random combinations of the \textit{reduced} gate set $\{X_{0},X_{1},Z_{0},Z_{1}, \textrm{cz}_{0,1} \circ Z_{0} Z_{1}\}$. In this scenario, the error rate at the circuit output for the uncoded scheme can be approximated analytically by
\begin{equation}
    \tilde{D}_{u} = \mathcal{E}_{\mathcal{P}} + \mathcal{E}_{M} = \frac{L}{5}\Big[6 \epsilon_{1} + \epsilon_{2}\Big] + 2P_{m} - P_{m}^{2}.
\end{equation}
Here the error rate for each circuit block $\mathcal{P}$ is taken to be the average gate error evaluated over the reduced gate set. This is defined by the gate error probabilities according to the physical gate count\footnote{For example, using the gate error probabilities according to the physical gate count in Table~\ref{table:epsilon}, the average uncoded error rate of the reduced gate set is $\mathcal{P} = \frac{6}{5} \epsilon_{1} + \frac{1}{5} \epsilon_{2}$. Inserting this into Eq.~\eqref{eqn:gate_error_model} gives $\mathcal{E}_{\mathcal{P}}= L(\frac{6}{5} \epsilon_{1} + \frac{1}{5} \epsilon_{2})$, when only considering the largest terms.} summarized in Table~\ref{table:epsilon}. The overall gate error rate $\mathcal{E}_{\mathcal{P}}$ is then determined by Eq.~\eqref{eqn:gate_error_model}. In addition, the measurement error $\mathcal{E}_{M}$ is defined by Eq.~\eqref{eqn:uncoded_Pe}. The parameters applied in Fig.~\ref{fig:bogota_RS_06} are approximated by the device's same specific calibration metrics provided by IBMQ \cite{IBMref} for the device within the calibration cycle the experiment was run in. These metrics are taken as general guide, but they must be applied with some caution \cite{proctor2017randomized}. Moreover, the fitting of the model to the experimental results does not represent an accurate calibration of the device noise, since the model is incomplete. For example, the parameter $P_{m}$ may encompass some state preparation error in this model, therefore it does not accurately represent the scale of measurement error in the device. 

Nevertheless, the uncoded model gives a reasonable approximation of the increase in error rate with the gate sequence length. Therefore, it may be reasonable to assume that two-qubit gates constitute the dominant source of the uncoded error, and therefore it may be deemed plausible that the fault tolerance criterion $D_{c}<D_{u}$ is satisfied by the post-selected scheme associated with $r<1$ for sequence lengths of $L>10$.

The $[4,2,2]$-encoded scheme operating without post-selection\footnote{The $[4,2,2]$-encoded scheme without post-selection is considered here to compare the error rate before and after error detection, as well as to evaluate the error model proposed.} and represented by $r=1$ includes the gate errors of the encoding circuit (without post-selection) as well as both the gate and measurement errors. Under these assumptions, and upon considering the largest terms, the analytical error rate of the output becomes:
\begin{align}
  \mathcal{E}_{E}+ \mathcal{E}_{\mathcal{P}} + \mathcal{E}_{M}  = \epsilon_{1} + 3 \epsilon_{2} + L\Big[\frac{12}{5} \epsilon_{1} + 2 \epsilon_{1}^{2}\Big]+4P_{m} - 6P_{m}^{2},
  \end{align}
which applies the same metrics as the uncoded scheme. In this equation, $\mathcal{E}_{\mathcal{P}}$ is derived by the same method as that used for the uncoded scheme. However, in this case $\mathcal{E}_{\mathcal{P}}$ is calculated using the average gate error of the coded logical gate set (when $r=1$). The physical gate count for the coded scheme is summarized in Table~\ref{table:epsilon}. The upper bound gives a reasonable approximation of the experimental results. However, it is clear that the model of the post-selected ($r<1$) scheme is overly optimistic for comparison with the results obtained from the \textit{Ibmq\_Bogota} device. The $[4,2,2]$-encoded scheme relying on post-selection is approximately characterized by:
\begin{equation}
  \tilde{D}_{c} = \frac{8\epsilon_{2}}{15} + L2 \epsilon_{1}^{2}+ 6P_{m}^{2},
\end{equation}
which is lower-bounded by the post-selected encoder error, namely by $\mathcal{E}_{E} \rightarrow 8\epsilon_{2}/15$, as described in Section~\ref{sec:encoder_error}. This error floor is owing to the residual two-qubit gate errors in the encoding circuit that cannot be detected during post-selection. However, this assumption is not consistent with the experimental results in Fig.~\ref{fig:bogota_RS_06}, which exhibit an error rate that is almost an order of magnitude higher than this, closer to $D_{c} \approx 0.07$.

It is not unexpected that the experimental results will deviate from this simple model, since we can assume that many parameters are required for accurately characterising the time-variant behaviour of the device during each consecutive circuit execution. Additionally, both temporal and spatial independence has been assumed for all circuit components, which represents a simplistic model of a real device. Nevertheless, since the error rate of the $[4,2,2]$-encoded gate sequence does not increase with the gate sequence length $L$, it may be surmised that the error of the encoding circuit outweighs that of the encoded gate sequence. In addition, the encoding circuit may contain more significant errors than just two-qubit gate errors. Let us consider this interpretation further.

However, this model excludes qubit preparation errors, which may occur before the circuit is activated, while initializing the qubit register. A qubit preparation error occurring before the encoding circuit will be proliferated by the subsequent CNOT gates to a logical error that cannot be detected in the post-selection phase implemented in this scheme. For example, an $X$ error imposed on $q_{2}$ before the encoding circuit seen in Fig.~\ref{fig:coded_ibm_fig} would result in the preparation of the $|\overline{01}\rangle$ state, rather than the intended $|\overline{00}\rangle$. If the preparation error $P_{p}$ was modelled as a single-parameter channel as described in Section~\ref{sec:single_param}, this error would contribute a term on the order of $\mathcal{O}(P_{p})$ to the encoded error rate, hence resulting in an excessive lower-bound according to $p$. In addition, the uncoded scheme will also have an error rate on the order of $\mathcal{O}(P_{p})$. Note that it is expected that this error would be discarded, if the full circuit of Fig.~\ref{fig:st_prep_422} is implemented.

This model also excludes detuning or gate-coherence errors. A simple calibration error can be thought of as an inaccurate rotation of the gate's output state, effectively imposing the same error each time the gate is activated. Since this error is systematic, it will rapidly escalate if the gate is used repeatedly. This raises the dilemma whether it could be mitigated by re-calibrating the gate rotation. Alternatively, it may be hypothesized that there is a random fluctuation in the gate output's rotation within a certain range and therefore re-calibration of the gate may only have a limited effect. Note that the average error rate of the Hadamard gate ($\epsilon_{1}$) does not represent the contribution of a gate-coherence error to the final error rate, because it is not encompassed by the Pauli error model considered here. See Appendix~\ref{sec:coherent} for a further explanation of errors that may not be accounted for in the error model considered here.

It is quite plausible that this affects the weighting of the superposition of the encoded state $|\overline{00}\rangle$, rather than influencing an individual qubit error detectable in post-selection. Therefore, when determining the statistical distance between the non-ideal experimental results and an ideal output distribution of quantifying the error rate, the weighting of the superposition in the experimental encoded state must be close to the ideal one, otherwise the difference of the two distributions would cause a high error floor. This may be straightforwardly resolved by user-calibrated gate pulses relying on a hybrid classical-quantum algorithm without the need for fully characterizing the circuit noise.

\subsection{Experiment 2: Single Gate}
\label{sec:exp2}

\begin{figure}  
    \centering 
    \includegraphics[width=0.5\textwidth]{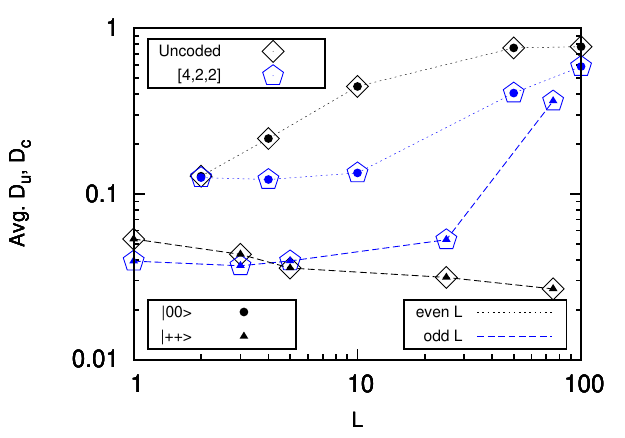}
    \caption[]%
    {{Experimental results based on the \textit{Ibmq\_Santiago} device characterizing the $[4,2,2]$-encoded $H_{0} H_{1} \circ SWAP_{0,1}$ gate repeated for the gate sequence lengths $L$. The final state at the circuit output is either $|00\rangle$ (1-dimensional) or $|{+}{+}\rangle$ (4-dimensional), which are both denoted by different symbols.}}    
    \label{fig:ibm_santiago_HHS_small}
\end{figure}

Fig.~\ref{fig:ibm_santiago_HHS_small} portrays the trace distance for the output states vs. the gate sequence length $L$, where the $H_{0} H_{1} \circ SWAP_{0,1}$ logical gates are applied $L$ times after the initialisation of the $|\overline{00}\rangle$ $[4,2,2]$-encoded state. When an \textit{even} number $L$ of gates is applied, the output state generated is
\begin{equation}
 [H_{0} H_{1} \circ SWAP_{0,1}]^{\otimes L}|00\rangle \rightarrow |00\rangle,
\end{equation}
since the effect of applying an even number of the same gate results in the identity operation\footnote{In the case where the gate sequence accumulates to an identity operation, the full gate sequence of length $L$ should still be implemented in the circuit. This is for ensuring that the experimental results show the effect of increasing the number of physical gates in the circuit with the sequence length. To do this in IBMQ experiments, the user should specifically select the compiler setting to avoid automatically simplifying the gate sequence to the most economical circuit.}. An \textit{odd} number $L$ of this gate, namely the  $H_{0} H_{1} \circ SWAP_{0,1}$ gate, will prepare the equi-probable 4-dimensional state
\begin{align}
    [H_{0} H_{1} \circ SWAP_{0,1}]^{\otimes L}|00\rangle \nonumber\\ \rightarrow \frac{1}{2}\Big( |00\rangle+|01\rangle+|10\rangle+|11\rangle \Big).
    \label{eqn:pp}
\end{align}
This is denoted by $|{+}{+}\rangle$ in Fig.~\ref{fig:ibm_santiago_HHS_small}. The figure shows that the circuits, which produce the output state $|00\rangle$ have an increasing trace distance vs. the gate sequence length $L$ and the circuits that generate the 4-dimensional state of Eq.~\eqref{eqn:pp} have a gently decreasing trace distance with the gate sequence length. 

Let us consider the trace distance for a 4-dimensional output state in the depolarizing channel, as described previously for the 1-dimensional output state in Section~\ref{sec:upperbound}. The circuit of Eq.~\eqref{eqn:++} generates a 4-dimensional ideal output state, given by
\begin{equation}
    p^{u}_{i} = \frac{1}{4} \quad \forall \quad i = \{ 00, 01, 10, 11\}.
    \label{eqn:4dideal}
\end{equation}
The trace distance between this state and the totally mixed state of Eq.~\eqref{eqn:uncoded_noisy} is
\begin{equation}
    D_{u} = \frac{1}{2}(4|p^{u}_{i}-\tilde{p}^{u}_{j}|) = 0.
    \label{eqn:4d}
\end{equation}
To interpret this effect in more detail, since the ideal state in Eq.~\ref{eqn:4dideal} and the totally corrupted \textit{worst case} noisy state in Eq.~\eqref{eqn:uncoded_noisy} are identical, the statistical trace distance between these states is zero. Therefore, when the ideal state is 4-dimensional, the error rate tends to $ D_{u} \rightarrow 0$, if the depolarizing noise affecting the experimental state obeys $\xi \rightarrow 1$. This explains the unexpected trend as to why the error rate may decrease even though the system error tends to become more prevalent. This trend is also seen for the logical 4-dimensional state in the encoded scheme.

Again, these trends are specific to the 4-dimensional output state. However, in general, the dimension of the output state will determine the upper bound of the trace distance. For example, consider a circuit where the 2-dimensional superposition state $(|10\rangle + |01\rangle)/\sqrt{2}$ is output, described by
\begin{equation}
    p^{u}_{01,10} = \frac{1}{2}, \quad p^{u}_{i} = 0 \quad \forall \quad i = \{ 00, 11\}.
\end{equation}
Let us employ the same reasoning to that in Eq.~\ref{eqn:4d}. When the depolarizing channel noise has its \textit{worst case} values associated with $\xi=1$, the measured experimental outcome is the totally corrupted state described by Eq.~\eqref{eqn:uncoded_noisy}. Then, in the noisiest scenario the uncoded error rate of Eq.\eqref{eqn:uncoded_D} becomes;
\begin{equation}
    D_{u} = \frac{1}{2}(2|p^{u}_{01,10}-\tilde{p}^{u}_{j}|+2|p^{u}_{i}-\tilde{p}^{u}_{j}|) = 0.5.
    \label{eqn:2d}
    \end{equation}
Therefore, when the ideal state is 2-dimensional, the error rate tends to $D_{u} \rightarrow 0.5$ as the depolarizing noise increases. Therefore, according to Eq.~\ref{eqn:4d} and Eq.~\ref{eqn:2d}, the dimension of the circuit output should be carefully considered, when assessing whether the fault tolerance criterion is satisfied.

Fig.~\ref{fig:ibm_santiago_HHS_small} demonstrates that the dimension of the output state will affect whether the circuit can or cannot satisfy the fault tolerance criterion. For example, the fault tolerance criterion of $D_{c}<D_{u}$ is only satisfied, when an even number of the $H_{0} H_{1} \circ SWAP_{0,1}$ gates are applied and the output state is 1-dimensional. The uncoded version of this gate has a SWAP operation, which is implemented with the aid of 3 CNOT gates, while its $[4,2,2]$-encoded version is implemented with the aid of single qubit gates, as shown in Table~\ref{table:gates}. Therefore, Fig.~\ref{fig:ibm_santiago_HHS_small} shows the trend that $D_{u} \rightarrow 0.75$ as $L \rightarrow 100$ only for even $L$, where the circuit output is $|00\rangle$. The corresponding encoded scheme has an error rate, which satisfies $D_{c}<D_{u}$, since the encoded circuit predominantly consists of single-qubit gates. However, when the output state is 4-dimensional, the reverse trend is observed. Since the uncoded scheme contains a large number of the noisiest gate, namely CNOT gates, the output state becomes more corrupted. However, this has the effect of mitigating the error rate\footnote{This is because the ideal output state is 4-dimensional, therefore it is expected that $D_{u}\rightarrow0$ when the experimental state sampled is totally corrupted.} as intimated in Eq.~\eqref{eqn:4d}. The logically equivalent state of the encoded scheme contains only single qubit gates, yet we have $D_{c}>D_{u}$. Therefore, for the fault tolerance criterion to be assessed, the gate sequences may be modified for ensuring that each circuit outputs a 1-dimensional state.

\subsection{Experiment 3: Full Gate Set}
\label{sec:other}

\begin{figure} 
\centering 
\includegraphics[width=0.5\textwidth]{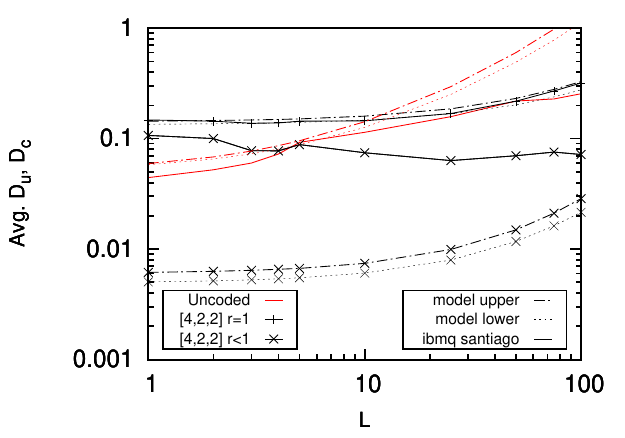}
\caption{Experimental results based on the \textit{Ibmq\_Santiago} device characterizing random sequences of the $[4,2,2]$-encoded gates and the corresponding uncoded gate for sequence lengths $L$. Model parameters: samples = 58, date= 14.02.2021, gate set = $[X_{0},X_{1},Z_{0},Z_{1}, \textrm{cz}_{0,1} \circ Z_{0}  Z_{1},H_{0} H_{1} \circ SWAP_{0,1}]$, $P_{m} = 0.025 $, $\epsilon_{1}:\epsilon_{2}$ = $1:20$ , $5 \times 10^{-3} < \epsilon_{1} < 6 \times 10^{-3}$.}
\label{fig:sant_FS_06}          
\end{figure}

\begin{figure} 
\centering 
\includegraphics[width=0.5\textwidth]{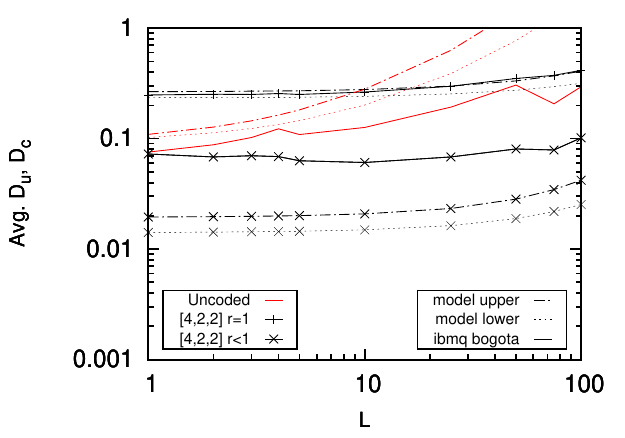}
\caption{Experimental results based on the \textit{Ibmq\_Bogota} device characterizing random sequences of the $[4,2,2]$-encoded gates and the corresponding uncoded gate for sequence lengths $L$. Model parameters: samples = 60, date= 15.04.2021, Gate set= $[X_{0},X_{1},Z_{0},Z_{1}, \textrm{cz}_{0,1} \circ Z_{0} Z_{1}, H_{0} H_{1} \circ SWAP_{0,1}]$, $P_{m} = 0.04 $, $\epsilon_{1}:\epsilon_{2}$ = $1:50$, $3 \times 10^{-3} < \epsilon_{1} < 5 \times 10^{-3}$.}
\label{fig:bogota_FS_06}              
\end{figure}   
                   
Fig.~\ref{fig:sant_FS_06} and Fig.~\ref{fig:bogota_FS_06} show the results of random gate sequences of the full $[4,2,2]$-encoded gate set $\{X_{0},X_{1},Z_{0},Z_{1}, \textrm{cz}_{0,1} \circ Z_{0}  Z_{1}, H_{0} H_{1} \circ SWAP_{0,1}\}$, which also includes the $H_{0} H_{1} \circ SWAP_{0,1}$ gate introduced in Section~\ref{sec:gates}. In contrast to the trends of the previous section, both figures show that the uncoded error rate does not increase as expected according to the model of Section~\ref{sec:ibmqresults}. This is due to the inclusion of the $H_{0} H_{1} \circ SWAP_{0,1}$ gate in the random gate sequence generated. As the sequence length increases, it is more common for an odd number of this gate to be included in the random gate sequence. In this case the final output state is a 4-dimensional equi-probable superposition, which results in a reduction of the total error rate. This is because the ideal state is indistinguishable from a totally mixed state and therefore the average states become the same, despite being produced from two very different scenarios  \cite{oreshkov2009distinguishability}. 

\section{Conclusion}
\begin{table}
\caption{Summary of results in Fig.~\ref{fig:bogota_RS_06}, Fig.~\ref{fig:sant_FS_06} and Fig.~\ref{fig:bogota_FS_06}. $D_{u}$ is the error rate for random gate sequences of length $L$ and $D_{c}$ is the corresponding $[4,2,2]$-encoded gate sequence with post-selection ($r<1$). }
\centering
\begin{tabular}{c|cc|cc}
\hline \hline \multicolumn{1}{l|}{} & \multicolumn{2}{c|}{$\mathbf{L=10}$}           & \multicolumn{2}{c}{$\mathbf{L=100}$} \\ 
\hline  
\textbf{Fig.} & $\mathbf{D_{u}}$ & $\mathbf{D_{c}}$ & $\mathbf{D_{u}}$ & $\mathbf{D_{c}}$ \\ 
\hline 
\hline   
\ref{fig:bogota_RS_06} & 0.08 & 0.06 & 0.34 & 0.09 \\
\ref{fig:sant_FS_06} & 0.11 & 0.07 & 0.25 & 0.07 \\
\ref{fig:bogota_FS_06} & 0.13 & 0.06 & 0.29 & 0.10 \\ 
\hline
\end{tabular}
\label{table:summary}
\end{table}

The results of Section~\ref{sec:other}, do not lead to consistent conclusions, when grouping together circuits with ideal output states of different dimensions upon using the trace distance for evaluating the fault tolerance criterion. Despite this, $[4,2,2]$-encoded gate sequences satisfy the fault tolerance criterion for the full gate set due to the inclusion of a larger number of the noisiest gates in the uncoded scheme. The encoded performance may be further improved, when aiming for mitigating either state preparation or gate-coherence errors by the post-selection mechanism. This may leave space for a combined classical and quantum machine learning approach, whereby the device errors are estimated and mitigated for circumventing the need for a comprehensive characterisation of the device. 

An accurate noise model that encompasses all sources of errors will require many parameters, especially for numerous qubits and long gate sequences considering a range of different coherent and incoherent errors. However, such a noise model may become excessively complex, in particular, when it encompasses unique correlated error patterns. In conclusion, the results of Fig.~\ref{fig:bogota_RS_06}, Fig.~\ref{fig:sant_FS_06} and Fig.~\ref{fig:bogota_FS_06} are summarized at a glance in Table~\ref{table:summary}.

\linespread{1.0}

\bibliographystyle{ieeetr}

\section{Appendix}

\subsection{The Depolarizing Channel}
\label{sec:depolarizing}

The depolarizing quantum channel may be viewed as a quantum-domain relative of a classical binary symmetric channel \cite{N&C}, where the qubit error can be either a bit-flip (X), a phase-flip (Z) or a combination of both (Y). Each error-events are equally likely, when it is assumed that the channel is symmetric (or unbiased). These errors can be thought of as the application of a Pauli operator to the qubit state. A single-qubit depolarizing channel is characterized by
\begin{equation}
    \mathcal{E}(\rho, p) = (1-p) \rho + \frac{p}{3}(X\rho X + Y\rho Y + Z\rho Z),
    \label{eqn:depo}
\end{equation}
where $\rho$ is the initial quantum state. The qubit remains intact with probability $(1-p)$ and it is depolarized with probability $p$, where each type of Pauli error occurs with probability $p/3$. If we substitute $p = \frac{3}{4} \xi$ in Eq.~\eqref{eqn:depo}, then we have:
\begin{equation}
    \mathcal{E}(\rho,\xi) = (1-\frac{3}{4} \xi) \rho + \frac{\xi}{4}(X\rho X + Y\rho Y + Z\rho Z),
    \label{eqn:depo2}
\end{equation}
which is equivalent to 
\begin{equation}
    \mathcal{E}(\rho, \xi) = \frac{I}{d}\xi + (1-\xi)\rho
    \label{eqn:depo3}
\end{equation}
for a $d$-dimensional quantum system, where $d=2$ for a single qubit state and $d=2^{n}$ for an $n$-qubit state. This has a slightly different interpretation from Eq.~\eqref{eqn:depo}. It can be interpreted by assuming that the initial state $\rho$ is replaced with the maximally mixed state $I/2$ with probability $\xi$ and left untouched with a probability $(1-\xi)$, for $p \leq \frac{3}{4}$. The totally mixed state describes the state of a system, completely corrupted by noise or 'totally randomized'. 

A $d$-dimensional quantum system that is completely mixed is described by
\begin{equation}
    \frac{I}{d} = \frac{1}{d}\sum_{i=1}^{d} |i\rangle \langle i|,
    \label{eqn:mixed2}
\end{equation}
regardless of its initial state $\rho$. This has the geometrical interpretation as the centre of the Bloch sphere, which gives rise to a measurement outcome distribution, where all possible measurement outcomes are equi-probable and the state of the qubit is not known.

The action of the depolarizing channel can be explained as follows.  In general, if there are $\mathcal{J}$ individual operators in the channel and $N$ qubits are sent over the channel, then there are $\mathcal{J}^{N} - 1$ channel operators excluding the operator associated with $N$ identity operators. For example, for $N=2$ qubits subjected to $\mathcal{J}=4$ operators $\{X,Y,Z, I\}$, there are $2^{4} -1 = 15$ operators excluding $II$. These are applied with a probability of $\epsilon/(\mathcal{J}^{N}-1)$, except in the case of no errors (i.e $II$), which occurs with a probability of $1-\epsilon$. 

\subsection{Coherent Error Insertion}
\label{sec:coherent}

\begin{figure}[htp]
    \centering
    \includegraphics[width=0.49\textwidth]{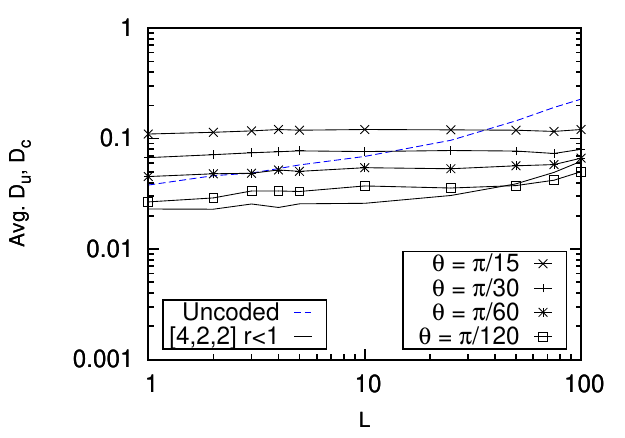}
    \caption{Experimental results from the \textit{Ibmq\_Jakarta} device showing random sequences of the $[4,2,2]$-encoded gates and of the corresponding uncoded gate for sequence lengths $L$. A rotation error $\theta$ is inserted after the Hadamard gate in the encoding circuit. 30 samples 21.06.2021}
    \label{fig:jakarta_c_2106}
\end{figure}

\begin{figure}[htp]
    \centering
    \includegraphics[width=0.49\textwidth]{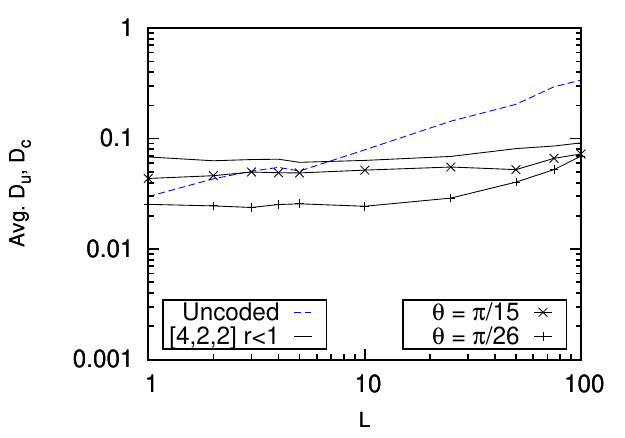}
    \caption{Experiment results from \textit{Ibmq\_Bogota} (same as Fig./~\ref{fig:bogota_RS_06}) with a rotation error $\theta$ inserted after the Hadamard gate in the encoding circuit. }
    \label{fig:santiago_c_2605}
\end{figure}

What are the most critical parameters when assessing a noise model? A tangible hypothesis is that rather than encountering a symmetric depolarizing noise channel, the device error is biased towards phase-flip errors ($Z$) rather than bit-flip errors ($X$). This is because the physical process of a phase-flip error is in a more direct interaction with the environment. However, there may be many parameters that accurately characterize a qubit that are not accounted for by the Pauli error model.

For example, a simple calibration error may be viewed as an over-rotation (or under-rotation) of a gate which has the same value each time the gate is activated. When this is systematic, the error is accumulated if the gate is used repeatedly and in theory can be resolved by improving the accuracy of the calibration of the gate rotation. Another type of gate error is constructed by dephasing errors. This varies randomly between each activation of the gate and it tends to be dependent on the time required to complete the associated operation. A leakage error is incoherent. This occurs when a qubit is relaxed from $|0\rangle$ to $|l\rangle$ at a probability of $p$. Crosstalk errors occur in two-qubit gates, which occur owing to the interactions between the target system connecting systems and they are also excluded from a traditional model of independent component errors \cite{gambetta2012characterization}.

\begin{figure}[htp]
    \centering
    \includegraphics[width=0.45\textwidth]{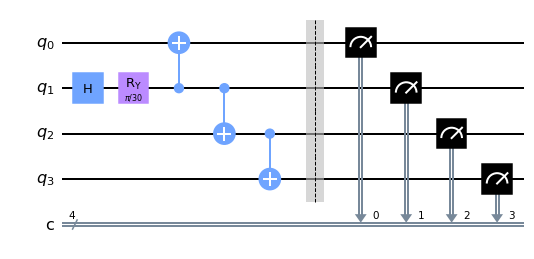}
    \caption{Coherent error insertion.}
    \label{fig:encoder_corr}
\end{figure}

Fig.~\ref{fig:jakarta_c_2106} and Fig.~\ref{fig:santiago_c_2605} show the results of inserting a rotation error $\theta$ in the encoding circuit before a random sequence of coded gates. This is demonstrated by the circuit shown in Fig.~\ref{fig:encoder_corr}. Fig.~\ref{fig:jakarta_c_2106} shows that as the rotation is increased, the error floor of the coded scheme is also increased. The opposite trend is seen in Fig.~\ref{fig:santiago_c_2605}. Therefore it may be concluded that this experiment is not robust to an over-rotation (or under-rotation) of the gate in the location shown in the circuit of Fig.~\ref{fig:encoder_corr}. Additionally, this error cannot be significantly mitigated by post-selection. 

\subsection{Other Encoded States}
\label{sec:other_encoded}
\begin{figure}[ht]
  \begin{minipage}[b]{0.45\linewidth}
    \includegraphics[width=.9\linewidth]{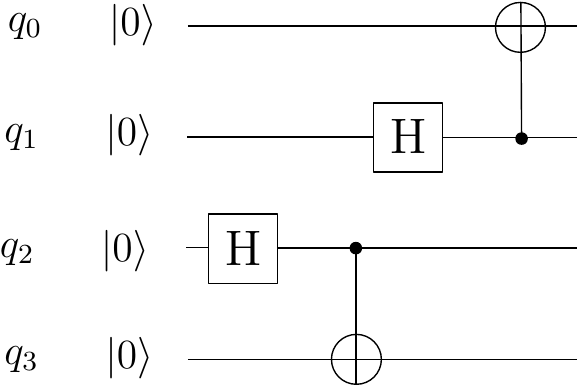} 
    \caption{Encoder for $|\overline{0+}\rangle$} 
    \label{fig:0p422}
  \end{minipage} 
  \begin{minipage}[b]{0.45\linewidth}
    \includegraphics[width=.9\linewidth]{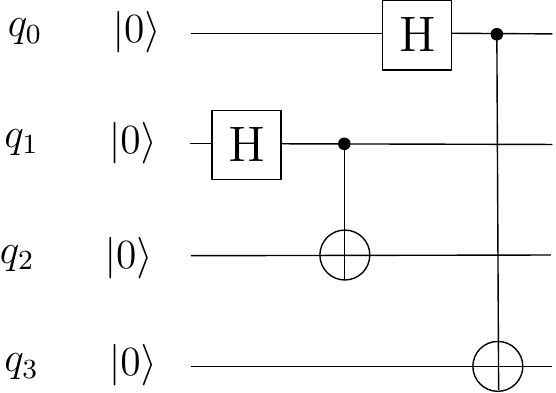} 
    \caption{Encoder for $|\overline{\phi^{+}}\rangle$} 
    \label{fig:psi0422}
  \end{minipage} 
\end{figure}

Apart from the circuit in Fig.~\ref{fig:st_prep_422}, other states that can also be directly prepared using the $[4,2,2]$ code, which are:
\begin{align}
    |\overline{0+}\rangle = (|\overline{00}\rangle +  |\overline{01}\rangle)\sqrt{2} \label{eqn:0p422}\\
    |\overline{\phi^{+}}\rangle = (|\overline{00}\rangle + |\overline{11}\rangle))\sqrt{2}. \label{eqn:phip422}
\end{align}

The circuit preparing the $|\overline{0+}\rangle$ state in Eq.~\eqref{eqn:0p422} is shown in Fig.~\ref{fig:0p422}. In this circuit there are two possible scenarios if either of the CNOT gates impose an error. Either it will result in a single qubit error, which can be picked up by post-selection, or the error will not affect the correctly prepared state. For example, an $XX$ Pauli error after the CNOT gate between $(q_{2}, q_{3})$ has no effect on the output distribution;
\begin{equation}
    |\overline{0+}\rangle = |0000\rangle + |1111\rangle + |1100\rangle + |0011\rangle)/2,
\end{equation}
since $|0000\rangle$ and $|0011\rangle$ are interchangeable. The circuit preparing the $|\overline{\phi^{+}}\rangle$ state shown in Fig.~\ref{fig:psi0422} is also fault-tolerant following a similar reasoning to that for Fig.~\ref{fig:0p422}.

\begin{IEEEbiography}[{\includegraphics[width=1in,height=1.25in,clip,keepaspectratio]{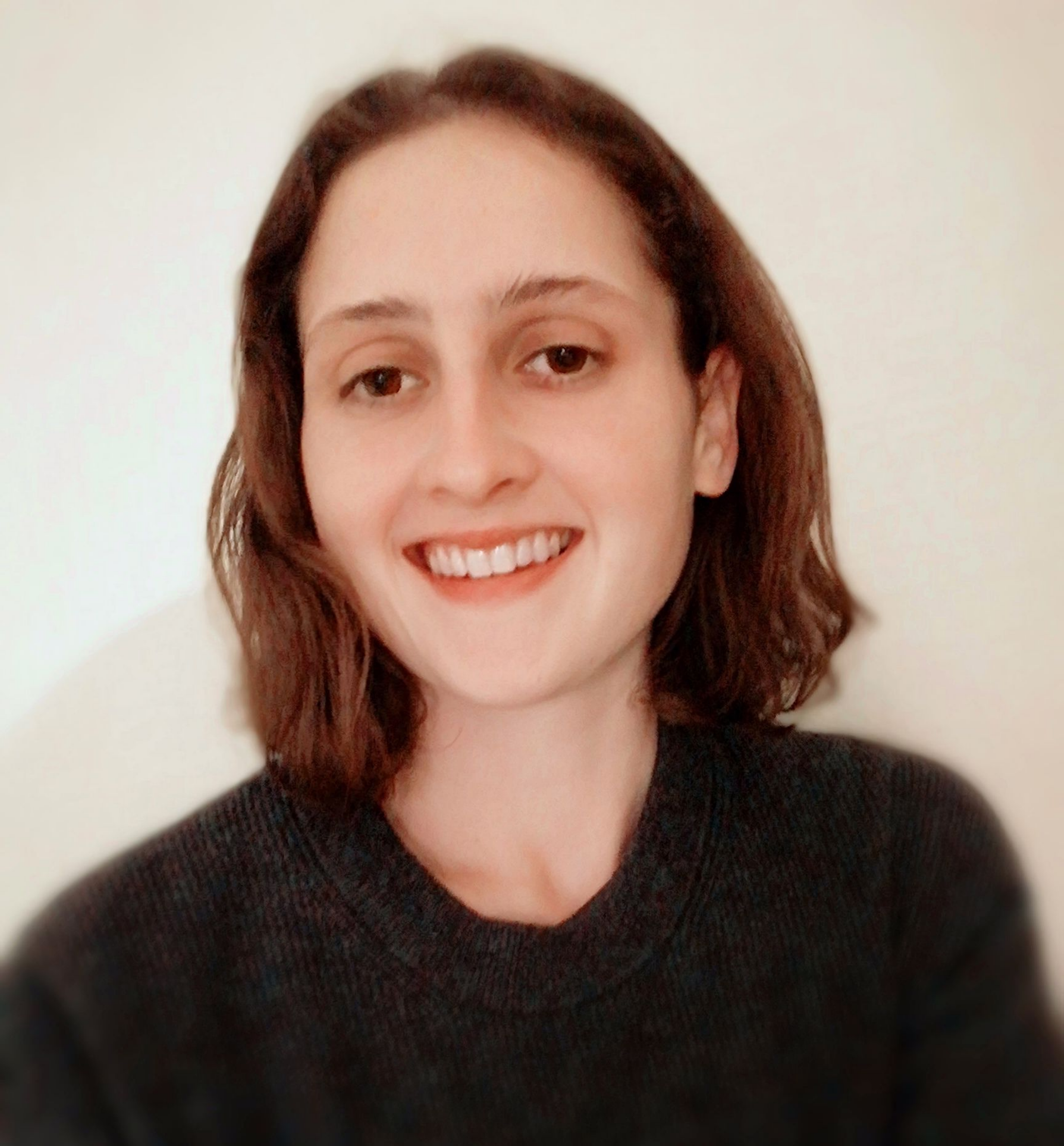}}]{\bf Rosie Cane} received the B.Sc. degree (Hons.) in physics and maths from the Open University, UK, in 2015, and the M.Sc. degree in wireless communications from the University of Southampton in 2017. She is currently pursuing the Ph.D. degree with Next-Generation Wireless within the School of Electronics and Computer Science, University of Southampton.
Her research interests include quantum computation and quantum information theory, quantum communications and quantum error correction codes.
\end{IEEEbiography}

\begin{IEEEbiography}[{\includegraphics[width=1in,height=1.25in,clip,keepaspectratio]{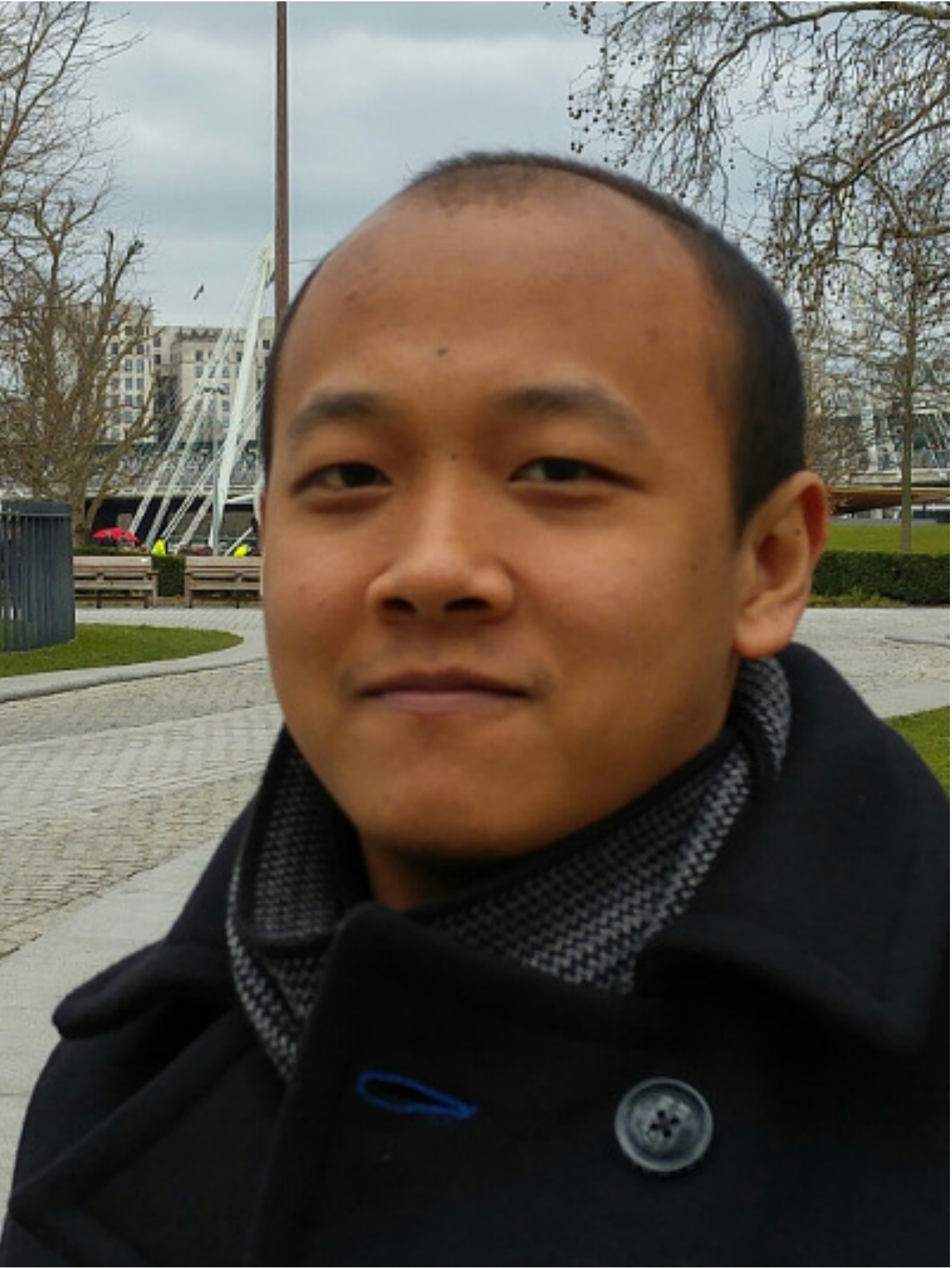}}] {\bf Daryus Chandra} received the M.Eng. degree in electrical engineering from Universitas Gadjah Mada, Indonesia, in 2014 and the Ph.D. degree in electronics and electrical engineering from University of Southampton, UK, in 2020. He was a research fellow with the Future Communications Laboratory, University of Naples Federico II, Italy. Currently, he is a research fellow with the Next-Generation Wireless Research Group, University of Southampton, UK. His research interests include classical and quantum error correction codes, quantum information, and quantum communications.
\end{IEEEbiography}

\begin{IEEEbiography}[{
\includegraphics[width=1in,height=1.25in,clip,keepaspectratio]{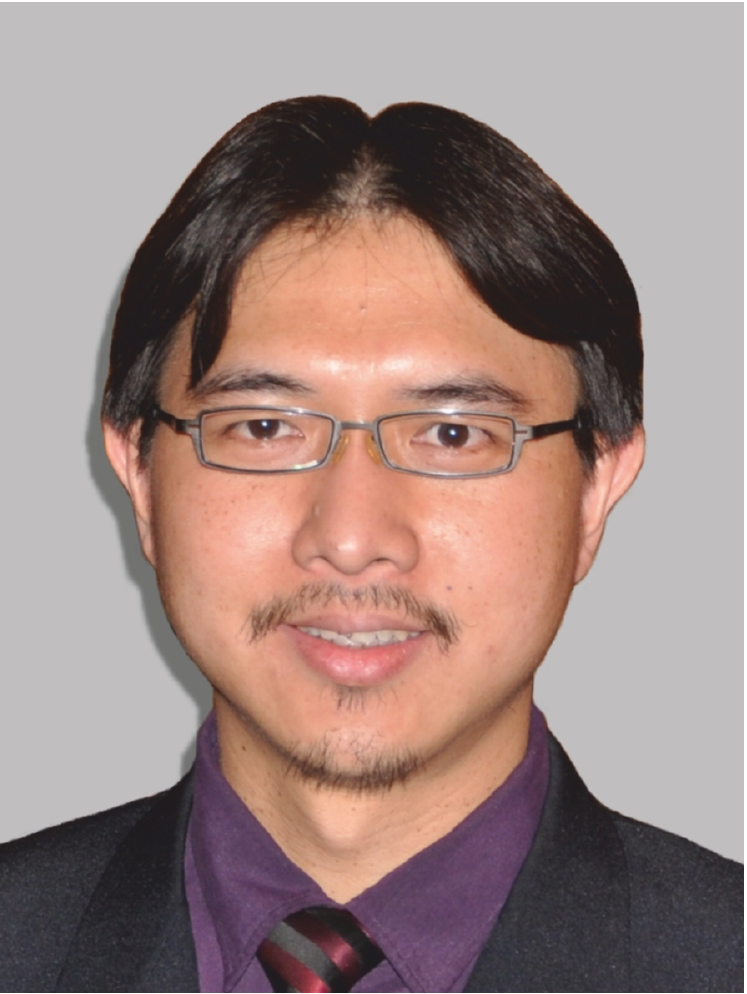}}]{\bf Soon Xin Ng} (S'99-M'03-SM'08) received the B.Eng. degree (First class) in electronic engineering and the Ph.D. degree in telecommunications from the University of Southampton, Southampton, U.K., in 1999 and 2002, respectively. From 2003 to 2006, he was a postdoctoral research fellow working on collaborative European research projects known as SCOUT, NEWCOM and PHOENIX. Since August 2006, he has been a member of academic staff in the School of Electronics and Computer Science, University of Southampton. He is involved in the OPTIMIX and CONCERTO European projects as well as the IU-ATC and UC4G projects. He is currently an Associate Professor in telecommunications at the University of Southampton.

His research interests include adaptive coded modulation, coded modulation, channel coding, space-time coding, joint source and channel coding, iterative detection, OFDM, MIMO, cooperative communications, distributed coding, quantum error correction codes and joint wireless-and-optical-fibre communications. He has published over 200 papers and co-authored two John Wiley/IEEE Press books in this field. He is a Senior Member of the IEEE, a Chartered Engineer and a Fellow of the Higher Education Academy in the UK.
\end{IEEEbiography}

\begin{IEEEbiography}
[{\includegraphics[width=1in,height=1.25in,clip,keepaspectratio]{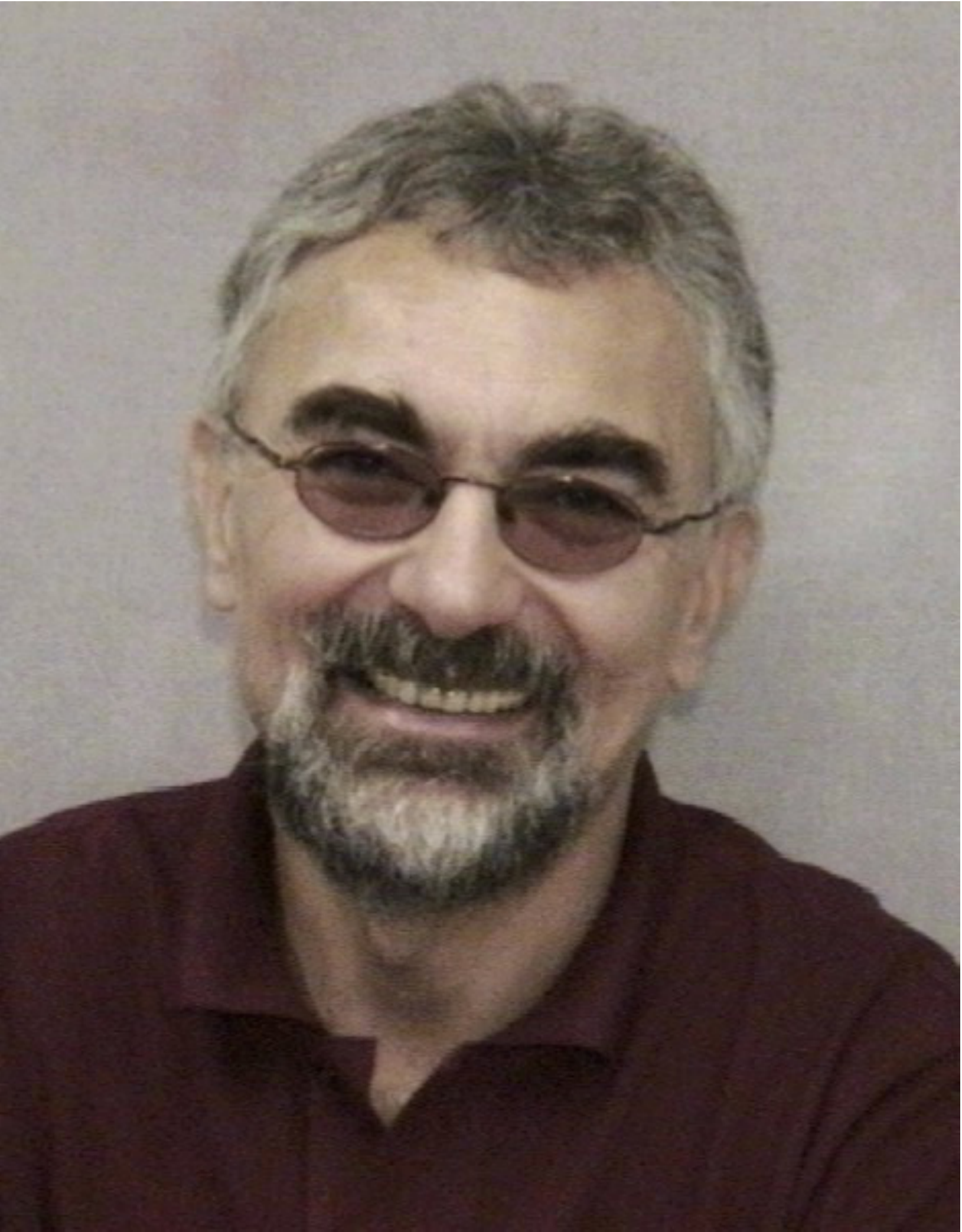}}] {\bf Lajos Hanzo} (M'91-SM'92-F'04) received his degree in electronics in 1976 and his doctorate in 1983. In 2009 he was awarded the honorary doctorate ``Doctor Honoris Causa'' by the Technical University of Budapest. During his 42-year career in telecommunications he has held various research and academic posts in Hungary, Germany and the UK. Since 1986 he has been with the School of Electronics and Computer Science, University of Southampton, UK, where he holds the chair in telecommunications. He has successfully supervised 119 PhD students, co-authored 18 John Wiley/IEEE Press books on mobile radio communications totalling in excess of 10~000 pages, published 1900+ research entries at IEEE Xplore, acted both as TPC and General Chair of IEEE conferences, presented keynote lectures and has been awarded a number of distinctions. Currently he is directing a 100-strong academic research team, working on a range of research projects in the field of wireless multimedia communications sponsored by industry, the Engineering and Physical Sciences Research Council (EPSRC) UK, the European Research Council’s Advanced Fellow Grant and the Royal Society’s Wolfson Research Merit Award. He is an enthusiastic supporter of industrial and academic liaison and he offers a range of industrial courses.

Lajos is a Fellow of the Royal Academy of Engineering, of the Institution of Engineering and Technology, and of the European Association for Signal Processing. He is also a Governor of the IEEE VTS. During 2008--2012 he was the Editor-in-Chief of the IEEE Press and a Chaired Professor also at Tsinghua University, Beijing. For further information on research in progress and associated publications please refer to \url{http://www.wireless.ecs.soton.ac.uk} .
\end{IEEEbiography}

\end{document}